\documentclass[useamsfonts]{pasj00}
\usepackage{natbib,rotating,amssymb,longtable}

\begin{document}
\SetRunningHead{Saurabh Sharma et al.}{NGC 281: Star formation}

\title{Multiwavelength Study of NGC 281 Region}

\author{Saurabh \textsc{Sharma}\altaffilmark{1,2,3} \thanks{saurabh@aries.res.in}, A. K. \textsc{Pandey}\altaffilmark{1}, J. C. \textsc{Pandey}\altaffilmark{1}, N. \textsc{Chauhan}\altaffilmark{8},  K. \textsc{Ogura}\altaffilmark{4}, D. K. \textsc{Ojha}\altaffilmark{5}, J. \textsc{Borrissova}\altaffilmark{2}, H. \textsc{Mito}\altaffilmark{6}, T. \textsc{Verdugo}\altaffilmark{2} and B. C. \textsc{Bhatt}\altaffilmark{7}}

\altaffiltext{1}{Aryabhatta Research Institute of Observational Sciences (ARIES), Manora Peak, Nainital, 263 129, India}
\altaffiltext{2}{Departamento de F\'isica y Astronom\'ia, Universidad de Valpara\'{\i}so, Ave. Gran Breta\~na 1111, Valpara\'iso, Chile}
\altaffiltext{3}{INAF-Osservatorio Astrofisico di Arcetri, Largo E. Fermi 5, 50125 Firenze, Italy}
\altaffiltext{4}{Kokugakuin University, Higashi, Shibuya-ku, Tokyo 150-8440, Japan}
\altaffiltext{5}{Tata Institute of Fundamental Research, Mumbai - 400 005, India}
\altaffiltext{6}{Kiso Observatory, School of Science, University of Tokyo, Mitake-mura, Kiso-gun, Nagano 397-0101, Japan}
\altaffiltext{7}{CREST, Indian Institute of Astrophysics, Hosakote 562 114, India}
\altaffiltext{8}{Institute of Astronomy, National Central University, Jhongli 32001, Taiwan}

\KeyWords{Galaxy: open clusters and associations: individual (IC 1590) - stars: formation - stars: luminosity function, mass function - stars:pre-main-sequence} 

\maketitle

\begin{abstract}
We present a multiwavelength study of the NGC 281 complex which contains the young cluster IC 1590 at the center, using deep wide-field optical $UBVI_c$ photometry, slitless spectroscopy along with archival data sets in the near-infrared (NIR) and X-ray. The extent of IC 1590 is estimated to be $\sim$6.5 pc. The cluster region shows a relatively small amount of differential reddening. 
The majority of the identified young stellar objects (YSOs) are low mass PMS stars having age  $<1-2$ Myr and mass 0.5-3.5 M$_{\odot}$. 
The slope ($\Gamma$) of the mass function for IC 1590, in the mass range $2 < M/M_\odot \le 54$, is found to be $-1.11\pm0.15$. 
The slope of the $K$-band luminosity function ($0.37\pm0.07$) is similar to the average value ($\sim$0.4) reported for young clusters.  The distribution of gas and dust obtained from the IRAS, CO and radio maps indicates clumpy structures around  the central cluster. The radial distribution of the young stellar objects, their ages,  $\Delta$($H-K$) NIR-excess, and  the fraction of classical T Tauri stars suggest triggered  star formation at the periphery of the cluster region.  However, deeper optical, NIR and MIR observations are needed to have a conclusive view of star formation scenario in the region.
The properties of the Class 0/I and Class II sources detected by using the {\it Spitzer} mid-infrared observations indicate that a majority of the Class II sources are X-ray emitting stars, whereas X-ray emission is absent from the Class 0/I sources.  The spatial distribution of Class 0/I and Class II sources reveals the presence of three sub-clusters in the NGC 281 West region. 
\end{abstract}

\section{Introduction}

H II regions have been studied quite extensively in recent years on account of their close association with star formation.  There seems to be two modes of star formation associated with H II regions depending on the initial density distribution of the natal molecular cloud. One is the cluster mode which gives birth to a rich open clusters and the other is the dispersed mode which forms only loose clusters or aggregates of stars. Presumably, the former takes place in centrally condensed, massive clouds, whereas the latter occurs in clumpy, dispersed clouds (see e.g., Ogura 2006). These clusters/aggregates of stars emerging from their natal clouds can be the laboratories to address some of the fundamental questions of star formation. Trends in their evolutionary states and spatial distribution can help distinguish between various star formation scenarios such as spontaneous or triggered star formation. Triggered star formation is a complex process and makes an interesting and important topic of star formation.  The formation of massive stars feeds energy back into the nearby environments, irradiating, heating and compressing the remain of the natal molecular cloud. This feedback can have either destructive or constructive effects, but it is not clear which dominates in a given cloud or overall in a galaxy. Many examples exist in our Galaxy as well as in other galaxies where spatial distributions of young stellar objects (YSOs) and their ages suggest for triggered star formation (see e.g. Walborn et al. 2002, Oye et al. 2005, Deharveng et al. 2005, Sharma et al. 2007, Chauhan et al. 2009).

The HII region NGC 281/Sh2-184 ($\alpha_{2000}=00^h52^m$, $\delta_{2000}$= +56$^\circ$ 34$^\prime$ or l=123$^\circ$.07, b= -6$^\circ$.31) is located at a relatively high Galactic latitude and has the centrally located cluster IC 1590 (Guetter \& Turner 1997, Henning et al. 1994).  The brightest member of IC 1590 is an O-type Trapezium-like system HD 5005, whose component stars HD 5005ab (unresolved), HD 5005c, and HD 5005d have spectral types of O6.5 V, O8 V, and O9 V, respectively (Walborn 1973; Abt 1986; Guetter \& Turner 1997). Despite many measurements, the distance estimates of NGC 281 varies from 2.0 kpc to 3.7 kpc (cf. Sato et al. 2008). Recently, using the VLBI observations of an associated H$_2$O maser source Sato et al. (2008) derived a trigonometric parallax of $0.355\pm0.030$ milli arcsec, corresponding to a distance of $2.81\pm0.24$ kpc.

The NGC 281 region provides an excellent laboratory for studying  in detail star formation through the interaction of high mass stars with their surrounding cloud. Of special interest in this region is the possibility of triggered star formation occurring on two different scales: the large-scale ($\sim$300 pc) supernovae-triggered formation of the first generation OB stars and their associated cluster (Megeath et al. 2002, 2003), and the subsequent, triggered  sequential and ongoing star formation on a smaller scale ($\sim$1-10 pc);  the latter is taking place in an adjoining molecular cloud (NGC 281 West) probably through an interaction with an HII region (the NGC 281 nebula) excited by the first generation OB stars (Elmegreen \& Lada 1978; Megeath \& Wilson 1997; Sato et al. 2008).

The southwestern quadrant of the NGC 281 nebula is obscured by the adjoining molecular cloud NGC 281 West. Ongoing star formation in NGC 281 West is indicated by the presence of H$_2$O maser emission and IRAS sources within this cloud near its clumpy interface between the HII region. This star formation may have been triggered by the interaction of the molecular cloud with the HII region (Elmegreen \& Lada 1978; Megeath \& Wilson 1997). The NGC 281 molecular cloud complex  was mapped both in $^{12}$CO (J=1-0) and $^{13}$CO (J=1-0)  emission lines by Lee \& Jung (2003). The central radial velocity of the NGC 281 West molecular cloud, $V_{LSR}$ = 31 kms$^{-1}$ (Lee \& Jung 2003) agrees well with that of the H$_2$O maser emission in the cloud (Sato et al. 2007). Megeath et al. (2002, 2003) suggested that this cloud complex was formed in a fragmenting super-bubble, which gave birth to the first generation OB stars, and these OB stars have then ionised the surrounding gas which subsequently triggered next generation star formation in the neighboring clouds (Sato et al. 2008). 

Though both low-mass and high-mass star-forming regions can be studied at variety of wavelengths ranging from radio waves to X-rays, however most of the present knowledge about the HII region/open cluster NGC 281 has been inferred from studies outside the optical region. Henning et al. (1994) made a multiwavelength studies of the  NGC 281/IC 1590 region including  Stro$\ddot m$gren photometry of the bright cluster stars. The first detailed $UBV$ CCD photometry of 279 stars for the cluster was published by  Guetter \& Turner (1997). Their photometry terminates at a magnitude limit that is marginally brighter than the expected brightness of pre-main sequence (PMS) and T Tauri stars in the embedded cluster region. Keeping above discussion in mind we feel that NGC 281 is an appropriate target for a deep and wide field optical/infrared photometry. In this paper, we present deep wide-field optical $UBVI_c$ data and slitless spectroscopy. We supplement them with archival data collected from the surveys such as {\it Chandra}, {\it Spitzer}, 2MASS, IRAS and NVSS (NRAO VLA Sky Survey). Our aim is to  understand the global scenario of star formation under the effects of massive stars  in the whole NGC 281/IC 1590 region. In Section 2, we describe our optical CCD photometric and slitless spectroscopic observations and briefly the data reduction. In Section 3, we discuss the archival data set used in the present study. In the ensuing sections, we present the results and  discuss star formation scenarios in the  NGC 281 region.

\section{Observations and data reduction}

\subsection{Optical photometry}

\begin{figure*}
\centering
\FigureFile(80mm,80mm){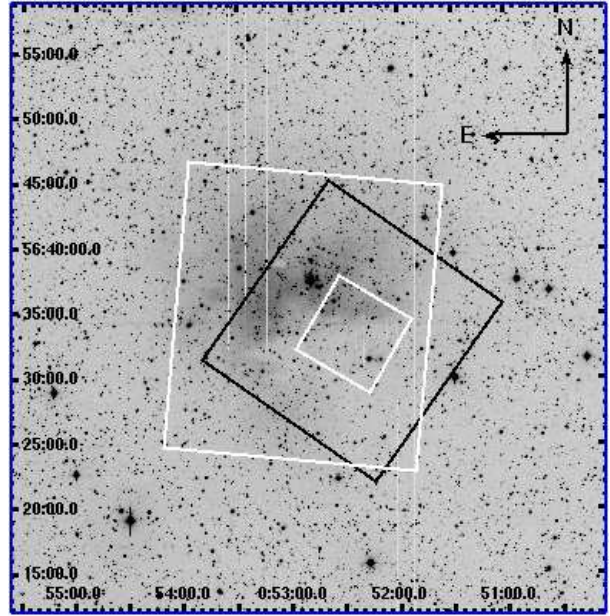}
\caption{$V$ band image of the NGC 281 region taken with the Kiso Schmidt.
The big and small white boxes are the areas covered by the H$\alpha$ and {\it Spitzer} observations and 
the black box represents the {\it Chandra} observations respectively.
The $X$ and $Y$ axes are in RA and Dec. in J2000.
\label{img} 
}
\end{figure*}

The CCD $UBVI_c$ observations of the NGC 281 region were obtained by using the 105-cm Schmidt telescope 
of the Kiso Observatory, Japan on November 21, 2004 and November 27, 2005. The CCD camera used a SITe 2048 $\times$ 2048  pixel$^2$
TK2048E chip having a pixel size of 24 $\mu$m. At the Schmidt focus (f/3.1), each pixel corresponds to 
1.5 arcsec and the entire chip covers a field of $\sim 50 \times 50$ arcmin$^2$ on the sky. 
The read-out noise and gain of the CCD are 23.2 $e^{-}$ and 3.4 $e^{-}/ADU$, respectively.
A number of short and deep exposure frames were taken. The average FWHM of star images was found to be $\sim$3 arcsec. 
The observed region is shown in Fig. \ref{img}.
The Kiso data were standardized by observing the cluster together with the standard stars in the SA 98 field
(Landolt 1992) on 07 January 2005 using the $2048\times 2048$ pixel$^2$ CCD camera 
mounted on the f/13 Cassegrain focus of the 104-cm Sampurnanand telescope of Aryabhatta Research Institute of 
Observational Sciences (ARIES), Nainital. In this set up, each pixel of the CCD corresponds to $\sim0.37$ arcsec 
and the entire chip covers a field of $\sim 13\times13$ arcmin$^2$ on the sky. To improve the signal to  noise ratio, this observation was carried out in the binning mode of $2\times2$ pixel.
The read-out noise and gain of the CCD are 5.3 $e^-$ and 10 $e^-$/ADU respectively. 
The FWHMs of the star images were $\sim2$ arcsec.  The log of these observations is given in Table \ref{log}. 

The CCD data frames were reduced by using computing facilities available at ARIES, Nainital.
Initial processing of the data frames was done
using the standard tasks available from IRAF\footnote{IRAF is distributed by National Optical Astronomy
Observatories, USA} and ESO-MIDAS\footnote{ ESO-MIDAS is developed and 
maintained by the European Southern Observatory.} data reduction packages. Photometry  of
cleaned frames was carried out by using the DAOPHOT-II software (Stetson 1987).
The PSF was obtained for each frame by using several uncontaminated
stars. Magnitudes obtained from different frames
were averaged. When brighter stars were saturated on deep exposure frames, their
magnitudes have been taken from short exposure frames.
We used the DAOGROW program in constructing of an aperture growth curve required for
determining the difference between aperture and profile fitting magnitudes.
Calibration of the instrumental magnitudes to those in the standard system was
done by using the procedures outlined by Stetson (1992).

 To translate the instrumental magnitudes to the standard magnitudes the following 
calibration equations, derived using a least-square linear regression, were used:\\

{\it  \small
\noindent
 $u= U + (7.004\pm0.004) -(0.005\pm0.006)(U-B) + (0.431\pm0.005)X$,
                                                                
\noindent                                                  
 $b= B + (4.742\pm0.005) -(0.035\pm0.004)(B-V) + (0.219\pm0.004)X$,
                                                                
\noindent                                                       
 $v= V + (4.298\pm0.002) -(0.038\pm0.002)(V-I) + (0.128\pm0.002)X$,
                                                                
\noindent                                                       
 $i= I_c + (4.701\pm0.004) -(0.059\pm0.003)(V-I) + (0.044\pm0.003)X$ \\

}

where $U,B,V$ and $I_c$ are the standard magnitudes and $u,b,v$ and $i$ are the instrumental aperture magnitudes normalized for 1 second of exposure time and $X$ is the airmass. We have ignored the second-order colour correction terms as they are generally small in comparison to other errors present in the photometric data reduction. The standard deviations of the standardization residuals, $\Delta$, between standard and transformed $V$ magnitude and $(U-B),(B-V)$ and $(V-I)$ colours of the standard stars are 0.006, 0.025, 0.015 and 0.015 mag, respectively. Short exposure data of the cluster region taken on the standardization nights, were standardized by using the above equations and coefficients. The standard magnitudes and colours of more than 50 stars obtained from these short exposures were further used to standardize the deep observations taken with Kiso Schmidt.  The standard deviations of the residual of secondary standards are of the order $\sim$0.02 mag.
The typical DAOPHOT errors in magnitude as a function of  corresponding magnitude in different pass-bands
for the Kiso Schmidt observations are found to increase with the magnitude and become large ($\ge$ 0.1 mag) for stars fainter than $V\simeq20$ mag. The measurements beyond this magnitude were not considered in the analysis.

\subsubsection{Completeness of the data}

To study luminosity functions (LFs)/ mass functions (MFs), it is very important to make necessary corrections in data sample to take 
into account the incompleteness that may occur for various reasons (e.g. crowding of the stars). 
We used the ADDSTAR routine of DAOPHOT II to determine the completeness factor (CF). 
The procedures have been outlined in detail in our earlier works (Pandey et al. 2001, 2005). 
 Briefly, the method consists of randomly adding artificial stars of known magnitude and position into the original frame. 
The frames are re-reduced using the same procedure used for the original frame.  
The ratio of the number of stars recovered to those added in each magnitude interval gives the CF as a function of magnitude.
In the case of optical CCD photometry, the incompleteness of the data increases with  magnitude as expected. The CF as a function of $V$ magnitude is given in Table \ref{completness}. Table \ref{completness} indicates that our optical data have a $95\%$ completeness at $V\sim16.5$ mag, which corresponds to a stellar mass of $\sim$ 2 M$_\odot$ for a PMS star having an age of $\sim$2 Myr (cf. Fig. \ref{cleaned}). 

\subsubsection{ Comparison with previous studies}

We have carried out a comparison of the present photometric data with those available in the literature. The difference $\Delta$ (literature - present data) as a function of $V$ magnitude 
is given in Table \ref{cmpt}. The comparison indicates that the present  $V$ mag and $B-V$ colour are 
in good agreement with the CCD  and photoelectric photometry by Guetter \& Turner (1997), whereas the $\Delta (U-B)$ shows a systematic variation with the $V$ magnitude
in the sense that the present $(U-B)$ colours become blue with increasing $V$ magnitude.

\subsection{Slitless grism spectroscopy}

Spectra of some PMS stars, specifically classical T-Tauri stars (CTTSs)
show emission lines, among which usually  H$\alpha$ is the strongest. Therefore, H$\alpha$ surveys
have often been used to identify PMS stars. 
We observed the NGC 281 region in the slitless mode with a grism as 
the dispersing element using the Himalayan Faint Object Spectrograph Camera (HFOSC) instrument 
during two observing runs on 10 October 2005 and 16 August 2006. 
This yields panoramic images where the
star images are replaced by their spectra. A combination of a `wide H$\alpha$' interference filter 
(6300 - 6740 \AA) and Grism 5 (resolution = 870) of HFOSC was used without any slit. 
The central $2K\times2K$ pixels of the $2K\times4K$ CCD
were used in the observations. The pixel size is 15 $\mu$m with an image scale of 0.297 arcsec pixel$^{-1}$.
The observed sky area is shown in Fig. \ref{img} as a large white box which was covered by four field-of-views of $\sim10 \times 10$ arcmin$^2$ each.
For each field-of-view we secured three spectroscopic frames of longer exposure with the grism in, and one direct frame of shorter exposure with the grism out for the identification purpose. 
The log of the observations is given in Table \ref{log}.
Emission line stars with enhancement over the continuum at the H$\alpha$ wavelength are visually identified. 

\section{Archival datasets}

\subsection{2MASS}

Near-infrared (NIR) $JHK_s$ data for point sources in the  NGC 281 region have been obtained from the Two Micron All Sky Survey (2MASS) Point Source Catalogue. 
The 2MASS data reported to be 99$\%$ complete up to $\sim 16, 15, 14.7$ mag in $J,H,K_S$ bands respectively\footnote{See http://www.ipac.caltech.edu/2mass/releases/allsky/doc/\\sec6\_5a1.html}.
To secure the photometric accuracy, we used only the photometric data with the quality flag ph-qual=AAA, which endorses a S/N$\ge10$ and photometric uncertainty $<$ 0.10 mag. The NIR data are used to identify the Classical T-Tauri stars (CTTSs) and Weak line T-Tauri stars (WTTSs) (cf. \S 4.3).

\subsection{$CHANDRA$ X-ray data}

Since YSOs are very strong X-ray emitters (as strong as log $L_X/L_{bol} \sim 10^{-3}$) and they can be detected behind column densities as large as N$_{HI}$ $\sim 10^{23}$ cm$^2$ (Linsky et al. 2007), X-ray imaging of star forming regions and young clusters is valuable for identifying these sources.  

\subsubsection{Observation}

{\it Chandra} observed the NGC 281 region on three occasions for 62.6 ks (Obs ID 5424, on 2005-11-10 @ 18:19:27 UT), 23.5 ks (Obs ID 7206, 2005-11-08 @ 13:41:46) and 13.1 ks (Obs ID 7205 @ 22:40:54 UT). The aim point of the array was $\alpha_{2000}$= $00^h52^m25^s.2$, $\delta_{2000}$ = $+56^\circ33^\prime47^{\prime\prime}.6$, and the satellite roll angle (i.e. the orientation of the CCD array relative to the north-south direction) was $225^\circ.5$ for all observations. The exposures were obtained in the very faint data mode with a 3.2 s frame time using the ACIS-I imaging array as the primary detector. ACIS-I consists of four front illuminated $1024\times1024$ CCDs with the pixel size of $\sim0.492$ arcsec  and the combined field of view of $\approx 16.9 \times 16.9$ arcmin$^2$. The S2 and S3 CCDs in ACIS-S were also enabled, however in the present study we used the ACIS-I data only. The detailed information on {\it Chandra}  and its instrumentation can be found in the {\it Chandra} Proposer's Guide (POG)\footnote{See http://asc.harvard.edu/proposer/POG}.
To detect sources, we have merged the event-list files of all observations.
The X-ray observed region is again shown in Fig. \ref{img} by a black box.

\subsubsection{Data reduction and Source detection}
We analyzed the data reprocessed by the {\it Chandra} X-Ray Center on 2006 April 5 (ASCDSVER 7.6.7.1).  The data were reduced by using the {\it Chandra} Interactive Analysis of Observations (CIAO; Fruscione et al. 2006) software (ver. 4.1; CALDB ver. 4.2). Light curves from the on-chip  background regions were inspected for large background fluctuations that might have resulted from solar flares, and none were found. We have filtered the data for the energy band 0.5 to 7.5 keV.  After filtering in energy, the time integrated background is 0.11 counts arcsec$^{-2}$.  Source detection was performed on the merge-event list by using the 1.7 keV exposure map with the PWDetect\footnote{See http://www.astropa.unipa.it/progetti\_ricerca/PWDetect/} code (Damiani et al. 1997), a wavelet-based source detection algorithm. The significance  threshold was set to 5 $\sigma$ so as to ensure a maximum one spurious source per field.  We have detected 379 sources, out of which 9 sources either fell on the unexposed areas of the CCD or were doubly detected.  This implies that a total of 370 X-ray sources were detected in the NGC 281 field. 
 An IDL-based program ACIS Extract (AE; Broos et al 2010) was used to extract the photons from each candidate source in a polygonal
region which closely matches with the local PSFs.
The source free regions around the source were considered as background.
AE provides the Poisson probability of not
being a source. We have not considered those sources which have the probability of being non-existence
$>0.01$. Sixteen such sources were found in the catalogue. After removing these sources the
catalogue consists of 354 X-ray sources. Further, the median-detected photon energy for the point
sources were determined by using the ACIS Extract software package.
We estimated the background AGN rate within the {\it Chandra} field of view using the {\it Chandra} Deep Field (Brandt et al. 2001). At the 0.5-2 keV limiting flux of $5.6\times 10^{-16}$ erg s$^{-1}$ cm$^{-2}$, we expect to find 79 to 127 background objects within the {\it Chandra} field of view.  
The optical, 2MASS and IRAC counterparts of the X-ray sources were searched within a match radius of 1 arcsec  and the data are given in Table \ref{Txray}. A sample of the table is given here, whereas the complete table is available in the electronic form only. Out of 354 X-ray sources, 193 and 90 sources 
have NIR and optical counterparts respectively. All the optical counterparts of X-ray sources have NIR counterparts also.
The location of X-ray sources in NIR colour-colour diagram has been used to identify the probable WTTSs/ Class III sources. The completeness of the X-ray data has not been estimated. Since we are using the WTTSs/ Class III sources to study the spatial distribution of these sources and to support the results obtained on the basis of rather complete optical and NIR data, the incompleteness of the X-ray data will not have any significant effect on the results presented in this study.

\subsection{{\it Spitzer} IRAC data}

The {\it Spitzer} mid-infrared (MIR) surveys have enabled detailed censuses of YSOs in star forming regions. The classification of young stars as protostellar Class I or more evolved Class II sources with optically thick discs  is best accomplished by using their broadband spectral energy distributions (SEDs)  (Muench et al. 2007).

We have used archived MIR data observed with Infrared Array Camera (IRAC). We obtained basic calibrated data (BCD) using the software Leopard. The exposure time of each BCD was 10.4 sec and for each mosaic, 72 BCDs have been used. Mosaicking was performed by using the MOPEX software provided by {\it Spitzer} Science Center (SSC). All of our mosaics were built at the native instrument resolution of 1.2 arcsec pixel$^{-1}$ with the standard BCDs. 
In order to avoid source confusion due to crowding, {\it PSF} photometry for all the sources was carried out. We used the {\it DAOPHOT} package available with the IRAF photometry routine to detect sources and to perform photometry in each IRAC band. The FWHM of every detection is measured and all detections with a FWHM $>$3.6 arcsec are considered resolved and removed. The detections are also examined visually in each band to remove non-stellar objects and false detections. The sources with photometric uncertainties $<0.2$ mag in each band were considered as good detections. A total of 347 sources were detected in the 3.6 and 4.5 $\mu m$ bands, whereas only 35 sources could be detected in all the four bands.

Aperture photometry for well isolated sources was done by using an aperture radius of 3.6 arcsec with a concentric sky annulus of the inner and outer radii of 3.6 and 8.4 arcsec, respectively. We adopted the zero-point magnitudes for the standard aperture radius (12 arcsec) and background annulus of (12-22.4 arcsec) of 19.670, 18.921, 16.855 and 17.394 in the 3.6, 4.5, 5.8 and 8.0 $\mu m$ bands, respectively. Aperture corrections were also made by using the values described in IRAC Data Handbook (Reach et al. 2006). The necessary aperture correction for the {\it PSF} photometry was then calculated from the selected isolated sources and were applied to the {\it PSF} magnitudes of all the sources. The 2MASS, optical and X-ray counterparts of the IRAC sources were searched for within a match radius of 1 arcsec. These counterparts are given in Table \ref{Tspit}. A sample of the table is given here, whereas the entire table is available in the electronic form. 
The completeness of the data in the 3.6, 4.5, 5.8 and 8.0 $\mu m$ bands having $S/N > 5$ (error $\le 0.2$ mag) is found to be $\sim$ 16.0, 15.5, 13.0 and 12.0 mag, respectively.

\subsection{IRAS}

The data from the IRAS survey in the four bands (12, 25, 60 and 100 $\mu$m) for the
NGC 281 region have been used to study the spatial
distribution of warm and cold interstellar dust.  One cold IRAS point sources is
identified in the cluster region and its details are given in Table \ref{iras}.

\section{Results}

\subsection{Structure of the cluster}

\subsubsection{Isodensity contours}

\begin{figure*}
\centering
\FigureFile(70mm,70mm){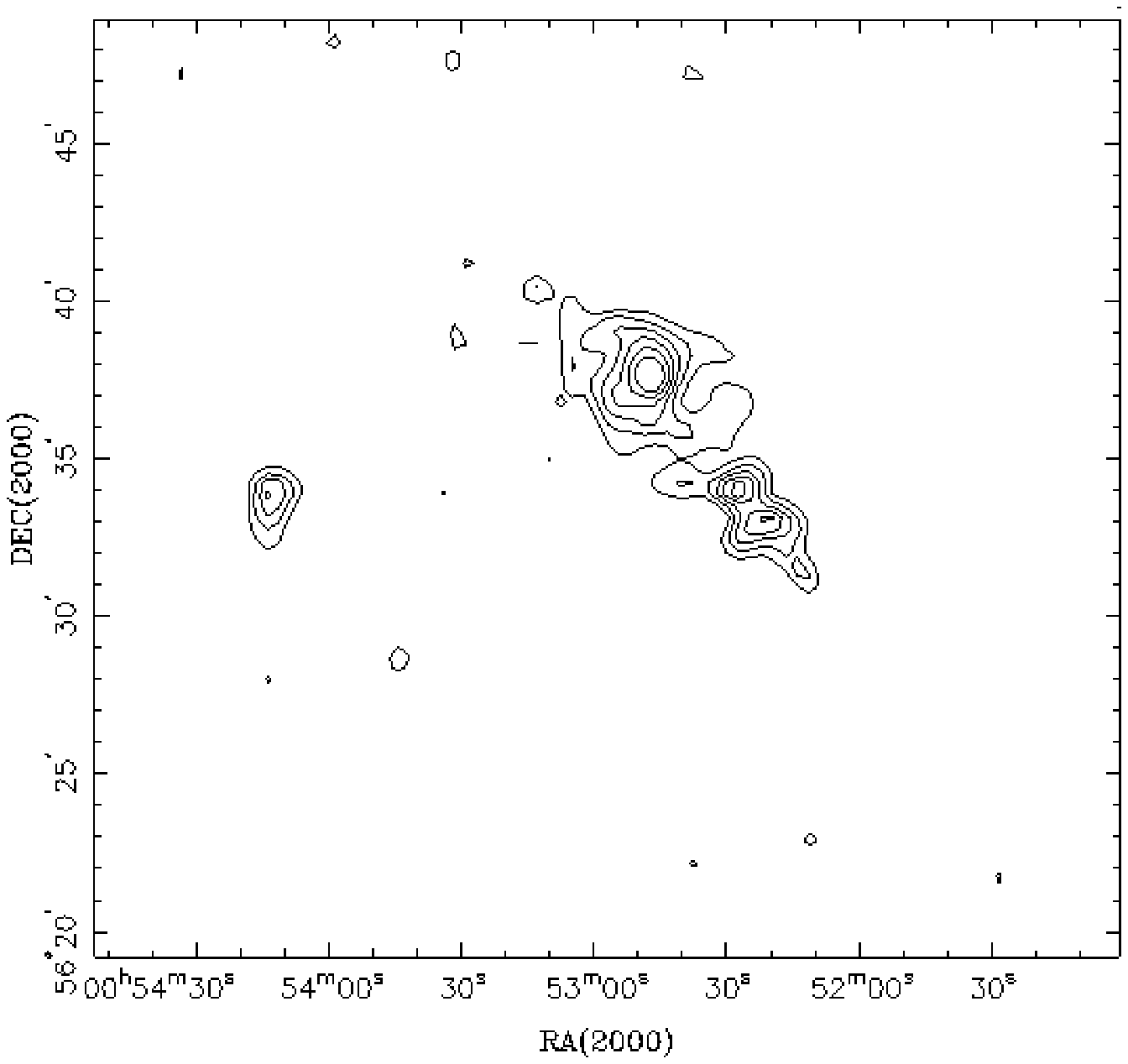}
\FigureFile(70mm,70mm){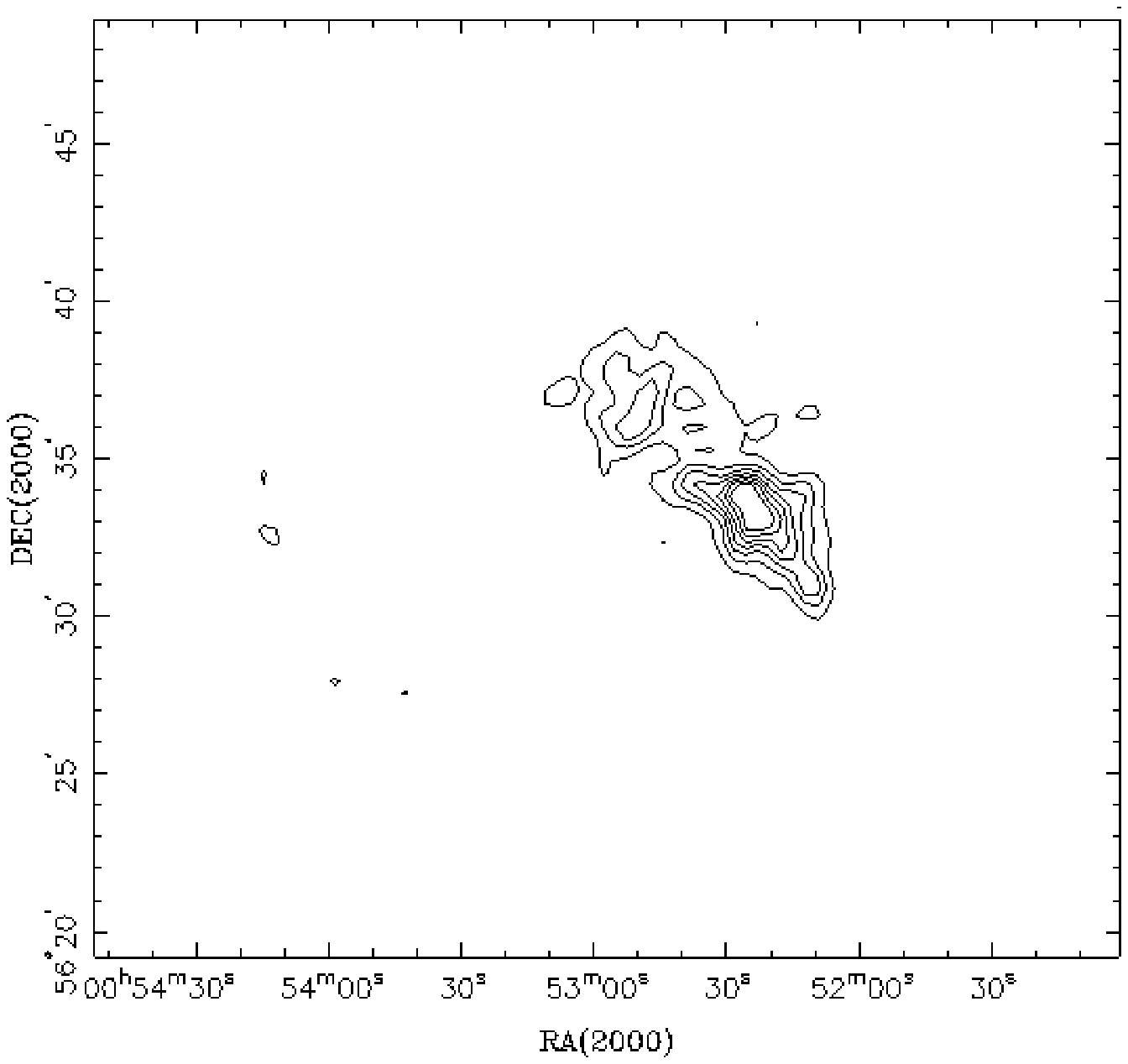}
\caption{ (left panel) Isodensity contours of the distribution of the 2MASS sources. The contours are plotted above 3 sigma levels with the step size of 5 stars/pc$^2$.  The lowest contour represents 17 stars/pc$^2$.
(right panel) Isodensity contours of the distribution of the identified YSOs. }
\label{iso}
\end{figure*}

Internal interaction due to two-body encounters among member
stars and external tidal forces due to the Galactic disc or giant molecular
clouds can significantly influence the morphology of clusters. However, in the
case of young clusters where dynamical relaxation is not important because of their
young age, the  stellar distribution can be considered as the initial state
of the cluster that should be governed by the star formation process in the parent
molecular cloud (Chen et al. 2004). To study the morphology of the  NGC 281 cluster,
we plotted isodensity contours using the 2MASS data  as well as table of the identified YSOs (cf. \S 4.3) in Fig. \ref{iso}.
The isodensity contours indicate an elongated morphology for the cluster.
It is interesting  to point out that sub-structures can be clearly seen in the 2MASS data (left panel of Fig. \ref{iso}) towards 
the south-west of the cluster as well as in the south-east.

\subsubsection{Radial stellar surface density profile}

To find out the extent and radial stellar density profile of IC 1590 we used 
the 2MASS data of $K\lesssim14.3$. First the cluster center was determined by using the stellar density
distribution in a 100 pixel wide strip along both the X and Y directions around an initially eye estimated center. The point of the maximum density obtained by fitting the Gaussian distribution is considered as the center of cluster. It is found to be  $\alpha_{(2000)} = 00^{\rm{h}} 52^{\rm{m}} 39^{\rm{s}}.5 \pm1.0^{\rm{s}}$, $\delta_{(2000)} = +56^{\circ} 37^{'} 46^{''}\pm15^{''} $.

\begin{figure*}
\centering
\FigureFile(70mm,80mm){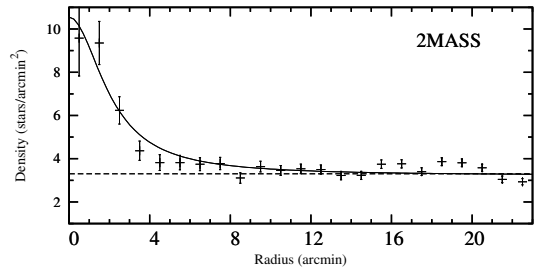}
\caption{Radial density profile of  the cluster using the 2MASS data.
The continuous  curve show  the least-square fit of the King (1962) profile to the observed data points.
The error bars represent $\pm\sqrt{N}$ errors. The dashed line indicate the density of field stars. }
\label{rdp}
\end{figure*}

To determine the radial surface density profile we assumed a spherical symmetry of stellar distribution and divided the cluster into a number
of concentric circles. The projected radial stellar density in each concentric annulus
was obtained by dividing the number of stars by
its area. The densities thus obtained are plotted in Fig. \ref{rdp}. The error bars are
derived by assuming that the number of stars in a concentric annulus follows
the Poisson statistics. The horizontal dashed line in the plot indicates the
density of contaminating field stars,  which is obtained from the reference region $\sim$ 15 arcmin
away toward the northwest from the cluster center  ($\alpha_{(2000)} = 00^{\rm{h}} 51^{\rm{m}} 16^{\rm{s}}.0$,
$\delta_{(2000)} = +56^{\circ} 46^{'} 45^{''}$).
The extent of the cluster $R_{cl}$ is defined as the projected radius from the density peak to the point at which the
radial density becomes constant and merges with the field star density.
$R_{cl}$ from the optical data as well as from the NIR 2MASS
data is estimated as $\sim8$ arcmin ($\sim$6.5 pc for a distance of 2.81 kpc).

The observed radial density profile of the cluster was parametrized by following the
approach of Kaluzny \& Udalski (1992). The projected radial density profile $\rho(r)$
is described as:
\begin{center}
$\rho (r) = {\rho_0\over \displaystyle{1+\left({r\over r_c}\right)^2}}$,
\end{center}
where the core radius $r_c$ is the radial distance at which the value of $\rho (r)$ becomes half of the central density $\rho_0$. 
The best fit obtained by the $\chi^2$ minimization technique is shown in Fig. \ref{rdp}.
Within the uncertainties the model reproduces well the observed radial density profile of IC 1590.
The core radius $r_{c}$ comes out to be $1.7\pm0.4$ arcmin ($1.4\pm0.3$ pc) 
and $2.0\pm0.3$ arcmin ($1.6\pm0.2$ pc) for the optical and 2MASS data, respectively.

\subsection{Interstellar extinction}

\subsubsection{\label{reddening} Reddening}

\begin{figure*}
\centering
\FigureFile(80mm,70mm){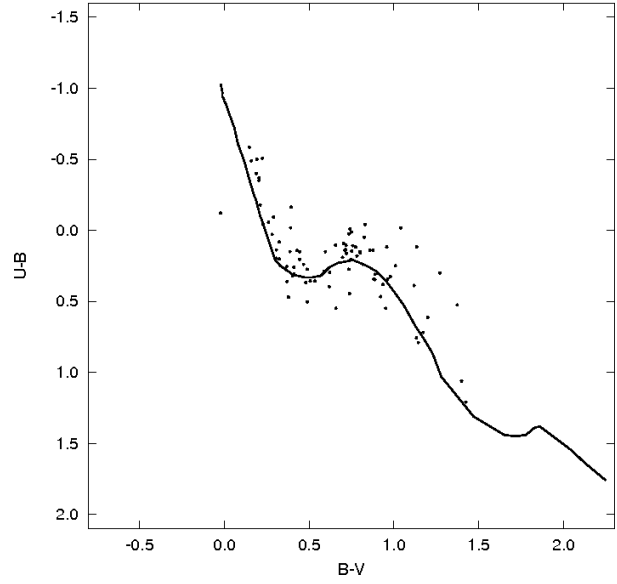}
\caption{The $(U-B)/(B-V)$ TCD for the stars lying within the cluster 
region ($r<5$ arcmin). The continuous curve represents the intrinsic MS by Schmidt-Kaler (1982) 
shifted along the reddening vector of 0.72 for $E(B-V)$ = 0.32 mag. }
\label{ccopt}
\end{figure*}

The interstellar extinction in the cluster region is studied by using the $(U-B)/(B-V)$ two-colour diagram (TCD) shown in Fig. \ref{ccopt} where zero-age-main-sequence (ZAMS) from Schmidt-Kaler (1982) is shifted along the normal reddening vector having a slope of $E(U-B)/E(B-V) = 0.72$. The distribution of stars shows a small amount of differential reddening ($E(B-V)\sim 0.2$ mag) in the region with  the minimum of $\sim 0.32$ mag which corresponds to the foreground extinction. 

The reddening for the individual star having a photometric error in the $V$ band $\sigma_V \le 0.1$ mag and of a spectral type earlier than $A0$ has also been  estimated by using the reddening free index $Q$ (Johnson \& Morgan 1953). 
Assuming the normal reddening slope we can construct the reddening-free index $Q = (U-B)-0.72 \times (B-V)$. For stars earlier than $A0$, the value of $Q$ will be less than 0.0. For main-sequence (MS) stars, the intrinsic $(B-V)_0$ colour and colour-excess can be obtained from the relation $(B-V)_0 = 0.332\times Q$ (Johnson 1966; Hillenbrand et al. 1993) and $E(B-V ) = (B-V ) - (B-V)_0$, respectively. The individual reddening of stars down to the $A0$ spectral type is found to vary in the range of $E(B-V )\sim$  0.32 - 0.52 mag, implying the presence of a small amount of differential reddening.

\subsubsection{Reddening law}

The extinction in star clusters arises due to two distinct sources; (i) the general interstellar medium (ISM) in the foreground of the cluster, and (ii) the localised dust associated with the cluster. While for the former component a value of $R = 3.1$ is well accepted (Wegner 1993; Lida et al. 1995; Winkler 1997), for the intra-cluster extinction the $R$ value varies from 2.42 (Tapia et al. 1991) to 4.9 (Pandey et al. 2000 and references therein) or even higher depending upon the conditions occurring in the region.

To study the nature of the extinction law in the IC 1590 region, we used TCDs as described in Pandey et al. (2000, 2003). The TCDs of the form of ($V-\lambda$) vs. ($B-V$), where $\lambda$ is one of the colour bands $R,I,J,H,K$ and $L$, provide an effective method for separating the influence of the possible abnormal extinction arising within intra-cluster regions having a peculiar distribution of dust sizes from that of the normal extinction produced by the diffuse interstellar medium (cf. Chini \& Wargau 1990, Pandey et al. 2000). The TCDs for the nearby reference region well away from IC 1590 (see \S 4.1.2) yield slopes of the distributions for $(V-I), (V-J), (V-H), (V-K)$ vs. $(B-V)$ as    $1.06\pm0.03, 1.97\pm0.05, 2.50\pm0.06,$ and $2.68\pm0.06$ respectively, manifesting a normal reddening law for the foreground interstellar matter (cf. Pandey et al. 2000).

\begin{figure*}
\centering
\FigureFile(150mm,50mm){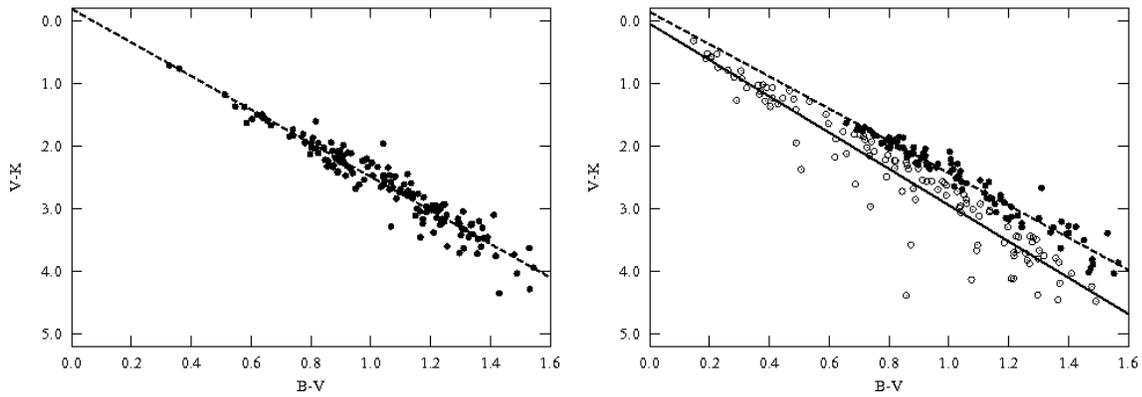}
\caption{ $(V-K)$ vs. $(B-V)$ TCDs for the nearby reference region ({\it left panel}) 
and for the cluster region ($r<R_{cl}$) ({\it right panel}. Open and filled circles represent probable cluster members  
and field stars with normal reddening, respectively. The continuous lines show the least square fits to the distributions of  the probable cluster members. The dashed lines show the fits to the distribution of field stars.
}
\label{tcd}
\end{figure*}

The $(V-K)$ vs. $(B-V)$ TCDs for the reference and cluster regions are shown in Fig. \ref{tcd}. The contamination due to field stars is apparent in the cluster region. We selected probable field stars having $(B-V) > 0.7$ visually, assuming that stars following the slope of the distribution of the reference region are contaminating foreground stars in the cluster region, and they are shown by filled circles.
The slopes of the distributions for the probable cluster members (open circles), $m_{cluster}$ are found to be $1.24\pm0.04, 2.16\pm0.06,2.76\pm0.08, 2.90\pm0.09$ for the $(V-I), (V-J), (V-H), (V-K)$ vs. $(B-V)$ TCDs respectively. The ratios ${E(V-\lambda)}\over {E(B-V)}$ and the ratio of the total-to-selective extinction in the cluster region, $R_{cluster}$, is then derived using the procedure given by Pandey et al. (2003). $R_{cluster}$ has turned out to be $3.5 \pm 0.3$.
From the photometry of bright cluster members in the $I$ and $K$ bands and a variable-extinction analysis of ZAMS members, Guetter \& Turner (1997)  also found a value of $R$ as $3.44\pm0.07$, which is comparable to ours.
Several studies have already pointed out anomalous reddening laws with high $R$ values in the vicinity of star forming regions (see e.g. Pandey et al. 2003 and references therein).
The higher than normal values of $R$ have been attributed to the presence of  larger dust grains. There is evidence that within dark clouds accretion of ice mantles on grains and their coagulation due to collision changes the size distribution towards larger dusts.

\subsection {Identification of YSOs}

\subsubsection {On the basis of $(J-H) / (H-K)$ TCD}

NIR imaging surveys are a powerful tool to detect YSOs in star forming regions. The locations of YSOs on $(J-H)/(H-K)$ two-colour diagrams (NIR TCDs) are determined to a large extent by their evolutionary state. Protostellar-like objects, CTTSs, weak-line T Tauri stars (WTTSs), Herbig Ae/Be stars, and classical Be stars tend to occupy different regions on NIR TCDs.

\begin{figure*}
\centering
\hbox{
\FigureFile(85mm,85mm){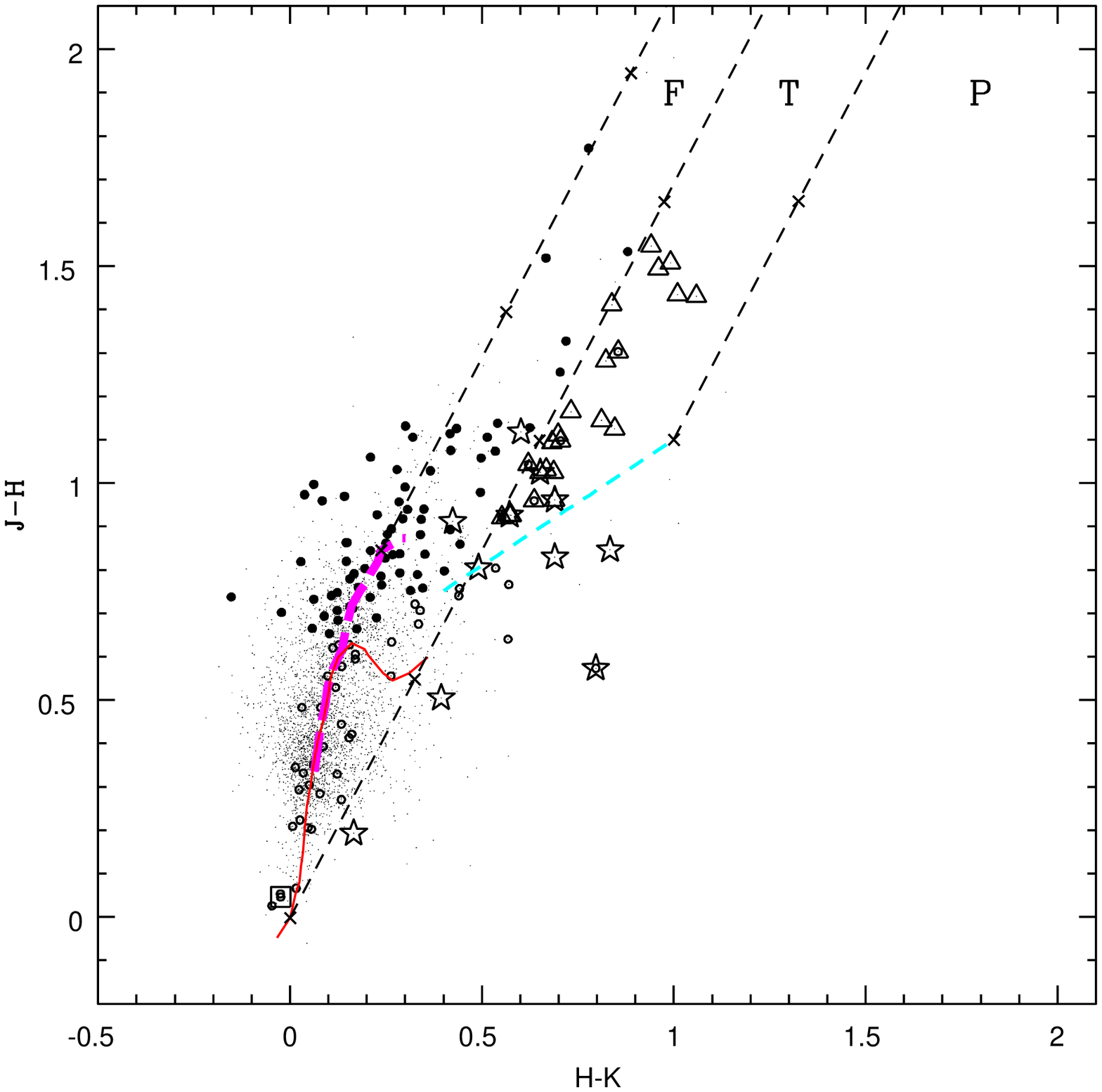}
\FigureFile(85mm,85mm){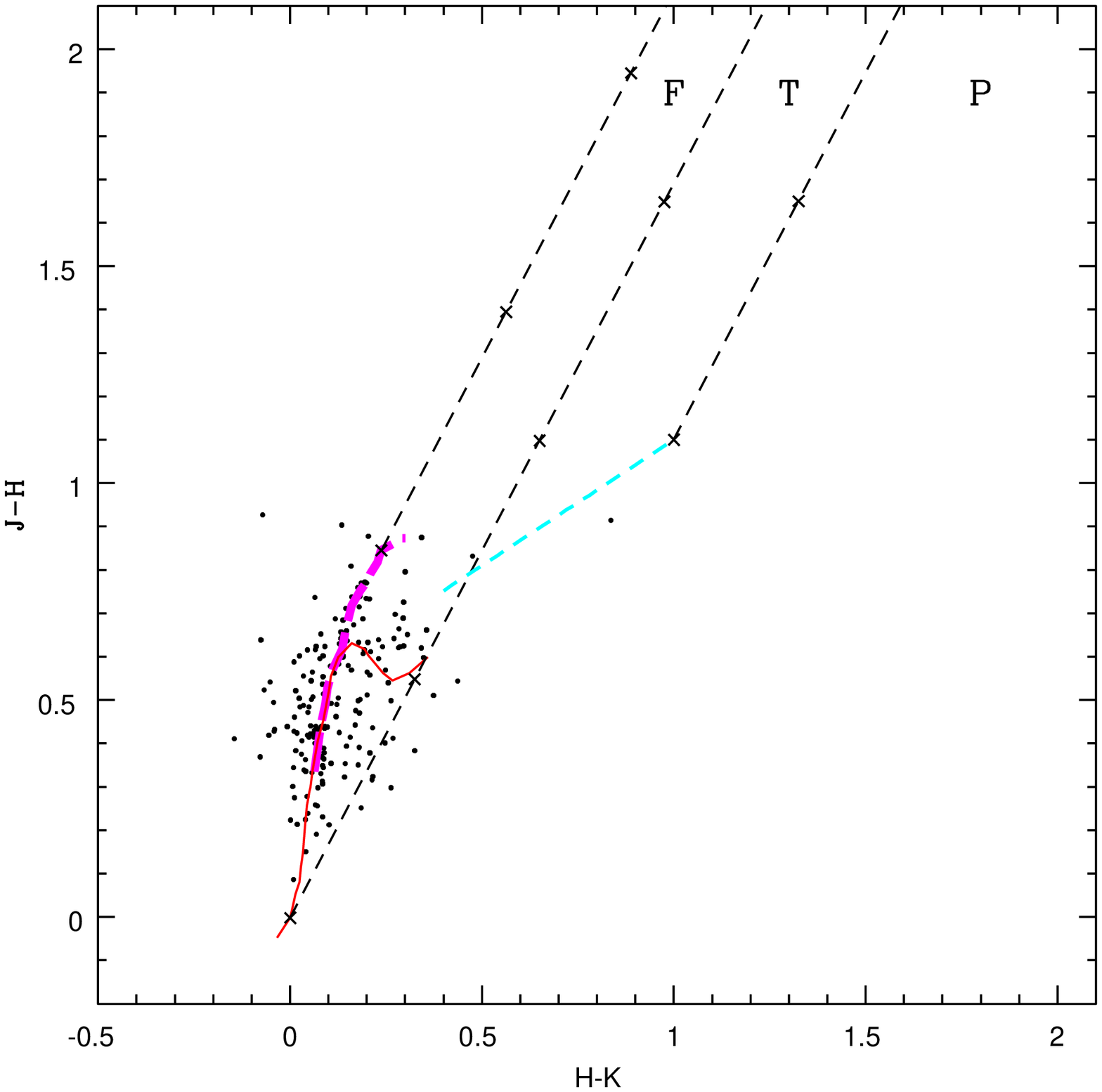}
}
\caption{Left panel: NIR  TCD of the  X-ray sources (open circles), $H\alpha$ emission (star symbols), NIR-excess sources (open triangles), probable WTTSs (filled circles) and O-type star (open square), having photometric errors less than 0.1 mag, in the NGC 281 region. Right panel: same as left panel but for all the  sources in the reference region  detected in the $JHK_s$ bands with the photometric errors less than 0.1 mag.
The sequences for dwarfs (solid curve) and giants (thick dashed curve) are taken from Bessell \& Brett (1988). The dotted line represents the loci of unreddened T Tauri stars (Meyer et al. 1997). Dashed straight lines represent the reddening vectors (see the text). The crosses on the dashed lines are separated by $A_V$ = 5 mag.
\label{nir-yso}
}
\end{figure*}

The NIR TCD   using the 2MASS data for all the sources lying in the NGC 281 region and  having photometric errors less than 0.1 magnitude is shown in the left panel of Fig. \ref{nir-yso}. All the 2MASS magnitudes and colours have been converted into the California Institute of Technology (CIT)  system.  The solid and thick dashed curves represent the unreddened MS and giant branch (Bessell \& Brett 1988) respectively.  The dotted line indicates the locus of unreddened CTTSs (Meyer et al. 1997).  All the curves and lines are also in the CIT system.  The parallel dashed lines are the reddening vectors drawn from the tip (spectral type M4) of the giant branch (``upper reddening line"), from the base (spectral type A0) of the MS branch (``middle reddening line") and from the tip of the intrinsic CTTS line (``lower reddening line").  The extinction ratios $A_J/A_V = 0.265, A_H/A_V = 0.155$ and $A_K/A_V=0.090$ have been taken from Cohen et al. (1981).  We classified sources according to three regions in this diagram (cf. Ojha et al. 2004a).  `F' sources are located between the upper and middle reddening lines and are considered to be either field stars (MS stars, giants) or Class III and Class II sources with small NIR-excesses. `T' sources are located between the middle and lower reddening lines. These sources are considered to be mostly CTTSs (or Class II objects) with large NIR-excesses. There may be an overlap of Herbig Ae/Be stars in the `T' region (Hillenbrand et al. 1992). `P' sources are those located in the region redward of the lower reddening line and are most likely Class I objects (protostar-like objects; Ojha et al. 2004a). It is worthwhile to mention also that Robitaille et al. (2006) have shown that there is a significant overlap between protostars and CTTSs.  The NIR TCD of the NGC 281 region (left panel of Fig. \ref{nir-yso}) indicates that a significant number of sources show $(H-K)$ excess and these are shown by open triangles. The sources having X-ray emission and H$\alpha$ emission are shown by circles (open and filled) and star symbols respectively. 
A comparison of the TCD of the NGC 281 region  with the NIR TCD of nearby reference region (right panel of Fig. \ref{nir-yso})  indicates that the sources in the NGC 281 region having X-ray emission and lying in the `F' region above the extension of the intrinsic CTTS locus as well as sources having $(J-H) \ge 0.6$ mag and lying to the left of the first (left-most) reddening vector (shown by filled circles) could be WTTSs/Class III sources. 
Here it is worthwhile to mention that some of the X-ray sources classified as WTTSs/ Class III sources, lying near the middle reddening vector could be CTTSs/ Class II sources.
The CTTSs and WTTSs identified in this section are listed in Table \ref{sltt}.

\subsubsection {On the basis of MIR data }

\begin{figure*}
\centering
\hbox{
\FigureFile(80mm,80mm){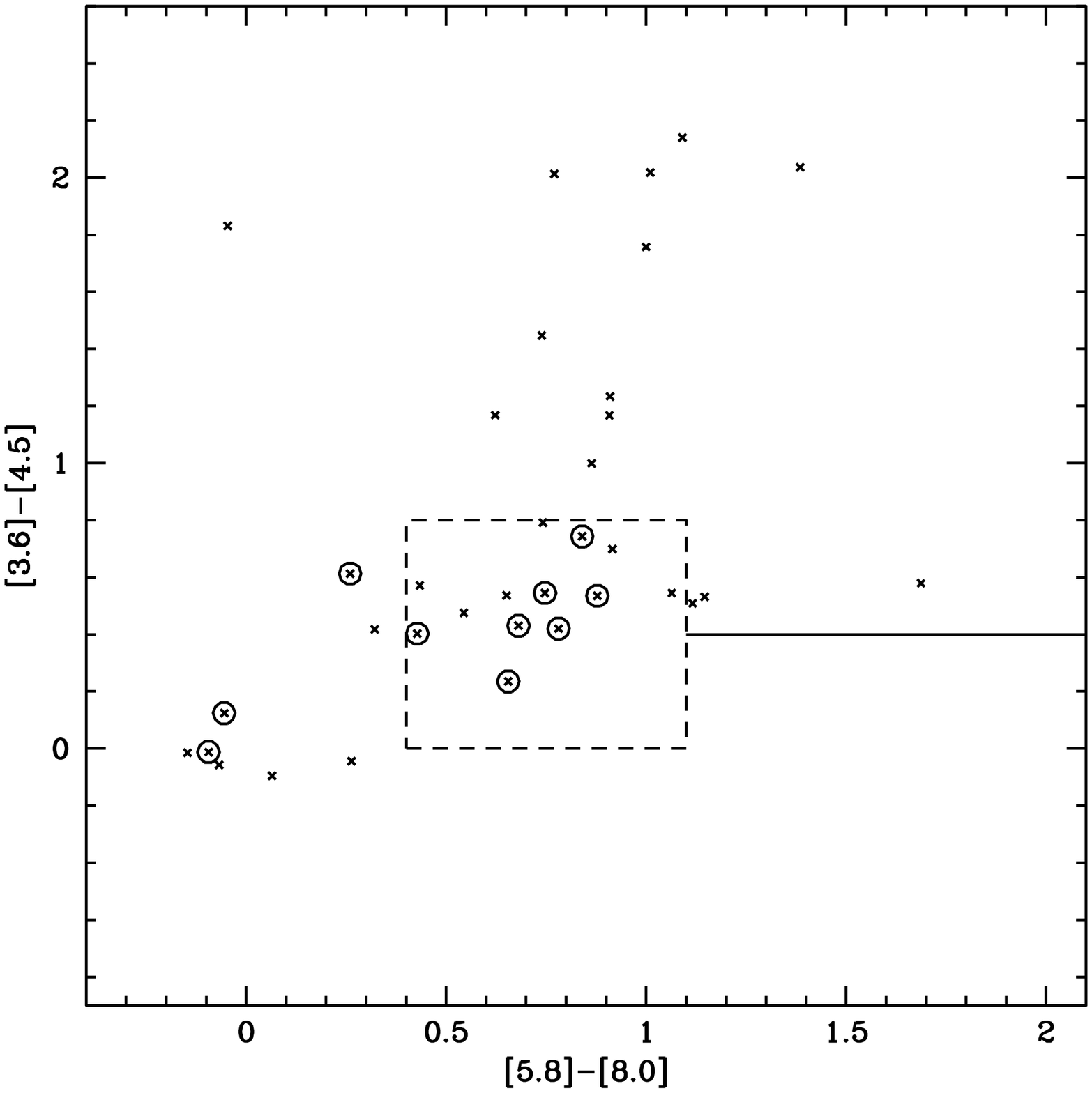}
\hspace{1.0cm}
\FigureFile(80mm,80mm){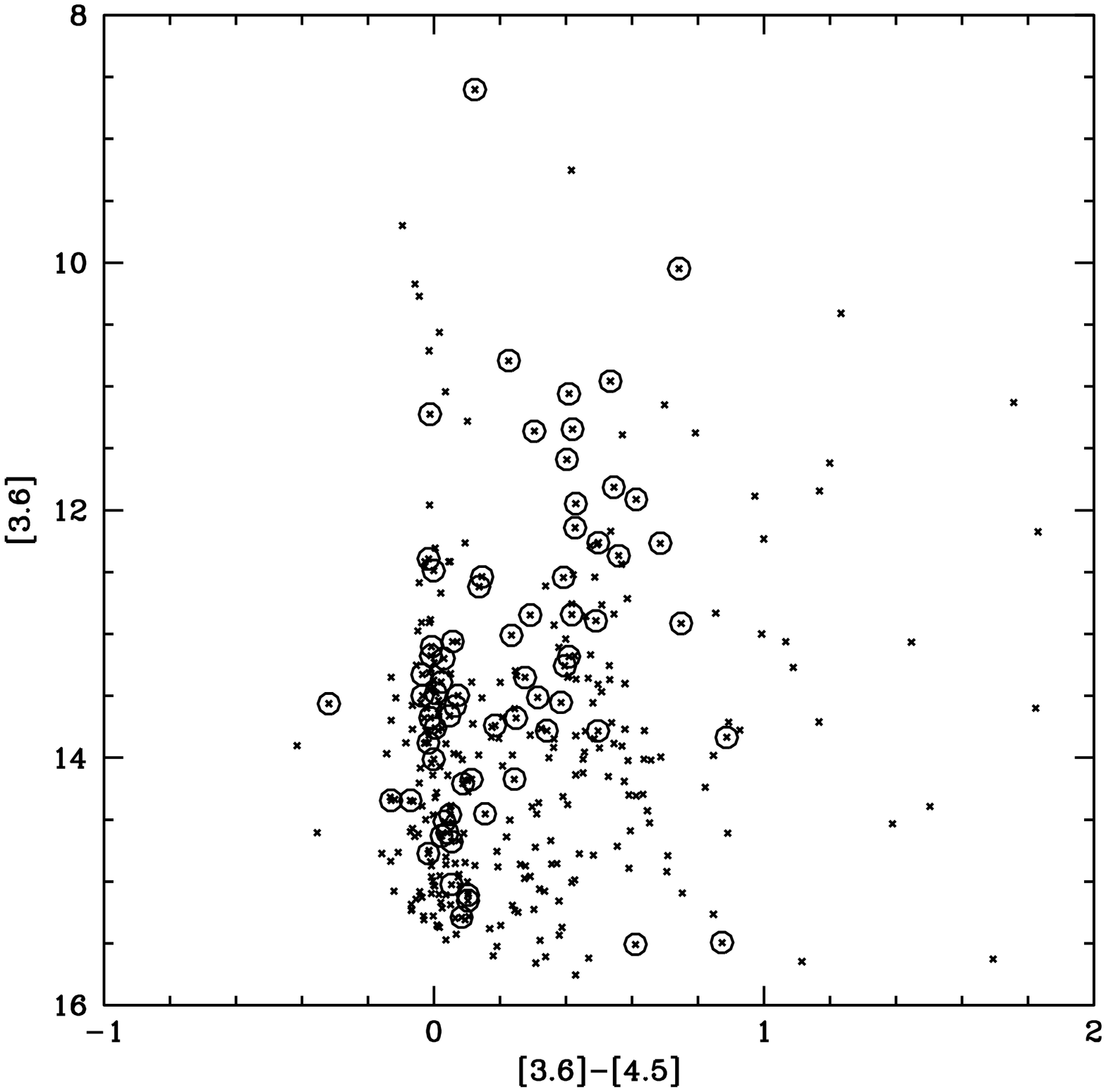}
}
\caption{ {\it (left)} IRAC MIR TCD of the detected sources. 
The sources lying within the box are Class II sources. The sources located
around [5.8]-[8.0] $\sim$ 0 and [3.6]-[4.5] $\sim$ 0 are field/Class III stars.
The sources with [3.6]-[4.5] $\ge$ 0.8 and/or [5.8]-[8.0] $\ge$ 1.1 represent Class 0/I
objects. The horizontal continuous line shows the adopted division between
Class I and Class I/II sources (see Megeath et al. 2004).
{\it (right)}  IRAC CMD for sources detected in the 3.6 and 4.5 $\mu m$ bands only.
Encircled sources represent objects with X-ray emission.  }
\label{spit}
\end{figure*}

The NGC 281 region  also has MIR observations through the {\it Spitzer Space Telescope} towards the south-west direction of the cluster.  Since young stars inside cloud clumps are  often deeply embedded, these MIR observations can provide a deeper insight into the embedded YSOs.   YSOs occupy distinct regions in the IRAC colour plane according to their nature; this makes MIR TCDs a very useful tool for the classification of YSOs.   Whitney et al. (2003) and Allen et al. (2004) presented independent model predictions for IRAC colours of various classes of YSOs. Fig. \ref{spit} {(left)} presents a [5.8]-[8.0] versus [3.6]-[4.5] TCD for the observed sources. The sources within the box represent the location of Class II objects (Allen et al. 2004; Megeath et al. 2004). The sources located around [5.8]-[8.0] = 0 and [3.6]-[4.5] = 0 are foreground/background stars as well as Class III objects. Sources with [3.6]-[4.5] $\ge$ 0.8 and/or [5.8]-[8.0] $\ge$ 1.1 have colours similar to those derived from models of protostellar objects with in-falling dusty envelopes (Allen et al.  2004). These are Class 0/I sources. Encircled objects represent sources with X-ray emission. A majority of the Class II objects have X-ray emission, whereas none of the Class 0/I sources show x-ray emission. 
It is found that four of the probable Class 0/I sources  identified on the basis of MIR data lie in the unexposed area of the detector (ACIS-I)
of the {\it Chandra} telescope, whereas one lies near the edge of detector ACIS-I of the {\it Chandra} telescope. 

The detection of Class 0/I and Class II sources in all the four IRAC bands is limited mainly by the lower sensitivity of the 5.0 and 8.0 $\mu m$ channels. Fig. \ref{spit} {(right panel)} shows the IRAC colour-magnitude diagram (CMD) for stars detected in the 3.6 and 4.5 $\mu m$ bands. Encircled objects represent sources with X-ray emission.  Stars having $ 0.35  \le  [3.6]-[4.5] \le 0.80$ mag  could be probable Class II stars, whereas stars having [3.6]-[4.5] $>$ 0.80 could be  Class 0/I sources. As can be seen, a majority of the Class II sources are X-ray emitting stars, whereas X-ray emission is mostly absent in  probable Class 0/I sources. One of the critical astrophysical questions is whether X-ray emission is present in Class 0/I sources at the very onset of star formation when collimated outflows begin (Getman et. al 2007). A few studies report detection of X-ray from Class 0/I protostars, whereas some studies reported that many bona-fide protostars are not detected in X-ray images (cf. Tsuboi et al. 2001, Hamaguchi et al. 2005, Getman et al. 2007). The non-detection of X-ray emission in Class 0/I sources is usually attributed to heavy obscuration instead of the intrinsic absence of X-ray  emission in protostars. 
 The  sources having  colours $ 0.35  \le  [3.6]-[4.5] \le 0.80$ mag and X-ray emission are considered as Class II sources and these are also  listed in Table \ref{sltt}.

\begin{figure*}
\centering
\FigureFile(160mm,160mm){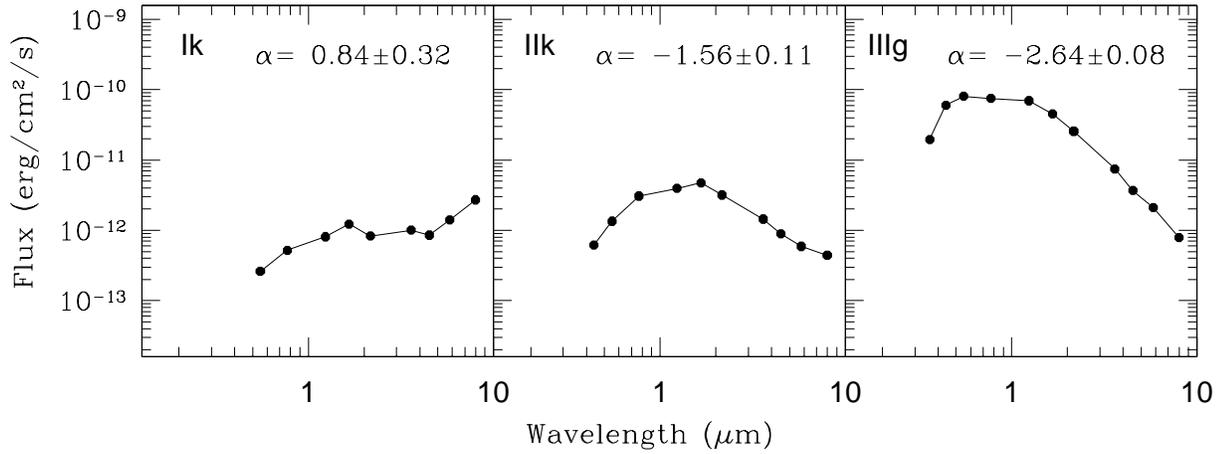}
\caption{ The sample SEDs obtained by using the optical, NIR and MIR observations.
Stars Ik, IIk and IIIg represent Class 0/1, Class II and Class III sources, respectively, classified according to the MIR TCDs 
(cf. \S 4.3.2 ) as given in Table \ref {spit}.  }
\label{seds}
\end{figure*}

To further elucidate the nature of the PMS sources, we derived SEDs for 35 sources (cf. Table \ref{data_all}) using the optical, NIR and MIR photometry.  In Fig. \ref{seds}, we show a sample of three SEDs for three different classes. To classify the evolutionary stage of YSOs using the SEDs, we adopted the classification scheme of Lada et al. (2006), which defines  the spectral class index  $\alpha$ = $d$log $\lambda(F_\lambda$)/$d$log($\lambda$). We computed the spectral class index $\alpha_{K-8 \mu m}$, which is the slope of the linear fit to the fluxes at the $K_s$ and IRAC 8 $\mu$m bands. Objects with  $\alpha_{K-8 \mu m} \geq +0.3$, $+0.3 >  \alpha_{K-8 \mu m} \geq -0.3$,  $-0.3 >  \alpha_{K-8 \mu m} \geq -1.8$ and  $-1.8 >  \alpha_{K-8 \mu m}$ are considered as Class I, Flat, Class II and Class III sources respectively.  The $\alpha$ indices obtained from the SEDs, in general, confirm the classification obtained from the MIR  TCD (cf. Fig. \ref{spit} (left)). However, the $\alpha_{K-8 \mu m}$ indices of two sources (Ih and Ij; Class I on the basis of MIR TCD) reveal that these must be Class II sources, whereas two seemingly Class II sources on the basis of the MIR TCD (IIa and IIc) appear to be Class I sources on the basis of the SEDs. The MIR TCD  classifies the two sources, namely IIIb and IIIh as Class III sources, whereas  $\alpha_{K-8 \mu m}$ indices classify them as Class II objects.

\begin{figure*}
\centering
\FigureFile(80mm,80mm){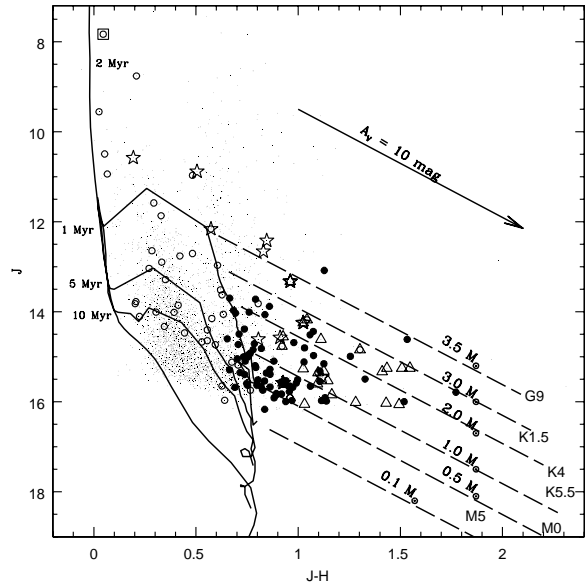}
\caption{$J/(J-H)$   CMD of  the X-ray sources (open circles), $H\alpha$ emission stars (star symbols), NIR-excess sources (open triangles), probable WTTSs (filled circles) and O-type star (open square),  having photometric errors less than 0.1 mag, in the NGC 281 region. The isochrone of 2 Myr (Z=0.02) and PMS isochrones of age 1, 5 and 10 Myr by Marigo et al. (2008) and Siess et al. (2000), respectively, corrected for a distance of 2.81 kpc and reddening $E(B-V)_{min}$ = 0.32 mag are also shown. 
The parallel slanting dashed lines denote loci of 1 Myr old PMS stars having masses in the range of 0.1 to 3.5 $M_\odot$ taken from Siess et al. (2000). } 
\label{cmdjhk}
\end{figure*}

Table \ref{sltt} provides a complete list of YSOs identified in the present study on the basis of H$\alpha$ emission, X-ray emission, NIR and MIR observations. The table contains 12 H$\alpha$ emission, 134 X-ray emission,  87  NIR excess (CTTSs) source and 118 WTTSs. The MIR data yield 25 and 61 Class I and Class II sources, respectively. The $J/(J-H)$ CMD (Fig. \ref{cmdjhk}) reveals that the identified YSOs are probably PMS stars of age $\lesssim$ 1 Myr. Majority of these stars have  masses between 0.5-3.5 M$_\odot$.

\subsection {Optical colour-magnitude diagram}

\begin{figure*}
\centering
\FigureFile(80mm,80mm){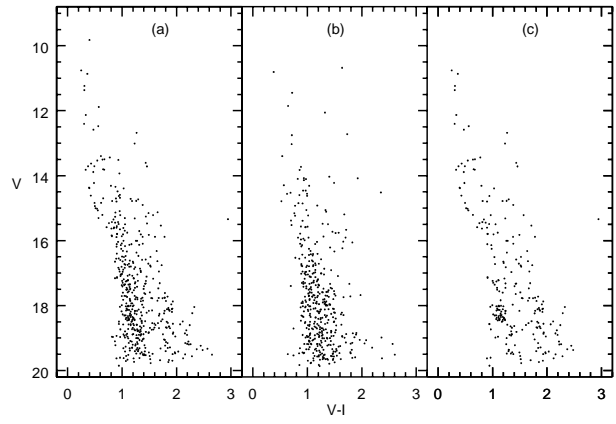}
\caption{$V/(V-I)$ CMD for (a) stars in the cluster region and (b) stars in the reference region. (c) is a statistically cleaned CMD.  }
\label{band}
\end{figure*}

The $V/(V-I)$ CMD for stars lying within the cluster region is shown in the left-hand panel of Fig. \ref{band}. A more or less well defined broad  MS presumably due to the variable reddening in the cluster region can be noticed down to $\sim$15 mag. The distribution of stars fainter than  $V \sim$ 15-16 mag deviates towards the red side of the MS indicating the presence of PMS stars in the cluster region. Contamination due to a field star population is also evident in the CMD. To study the LF/MF of the cluster, it is necessary to remove field star contamination from the sample of stars in the cluster region because PMS member stars and dwarf foreground stars both occupy similar positions above the ZAMS in the CMD. In the absence of  proper motion data, we used a statistical method to estimate the number of probable member stars in the cluster region. We again utilize the reference region towards the north-west (cf. \S 4.1.2); it has the same area as that of the cluster region. The middle panel of Fig. \ref{band} shows its $V/(V-I)$ CMD.

 To remove contamination due to field stars, we statistically subtracted their contribution from the CMD of the cluster region using the following procedure. The CMDs of the cluster as well as of the reference region were divided  into grids of $\Delta V=1$ mag by $\Delta (V-I) = 0.4$  mag. The number of stars in each grid of the CMDs were counted. After applying the completeness  corrections  using the CF (cf. Table \ref{completness}) to both the data samples, the probable number of cluster members in each grid were estimated by subtracting the corrected reference star counts from the corrected counts in the cluster region. The estimated numbers of contaminating field stars were removed from the cluster CMD in the following manner.  For a randomly selected star in the CMD of the reference region, the nearest star in the cluster CMD within $V\pm0.25$ and $(V-I) \pm 0.125$ of the field was removed.
Although the statistically cleaned $V/(V-I)$ CMD of the cluster region shown in Fig. \ref{band}c  clearly shows the presence of PMS stars in the cluster, however the contamination due to field stars at $V \gtrsim 17$ mag and $(V-I) \sim 1.2 $ mag can still be seen. This field population could be due to the background population as discussed by Pandey et al. (2006).

\begin{figure*}
\centering
\FigureFile(80mm,80mm){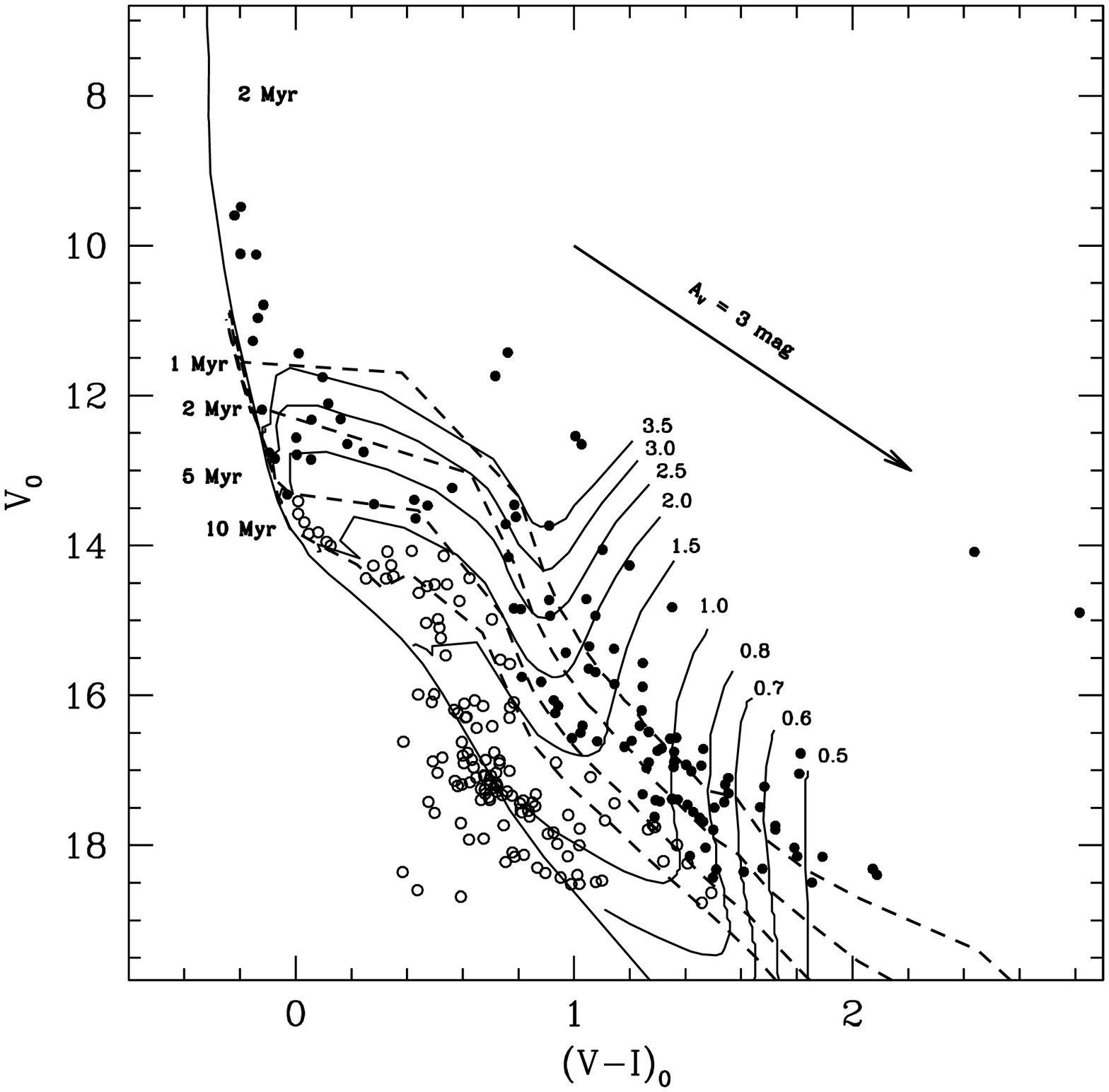}
\FigureFile(80mm,80mm){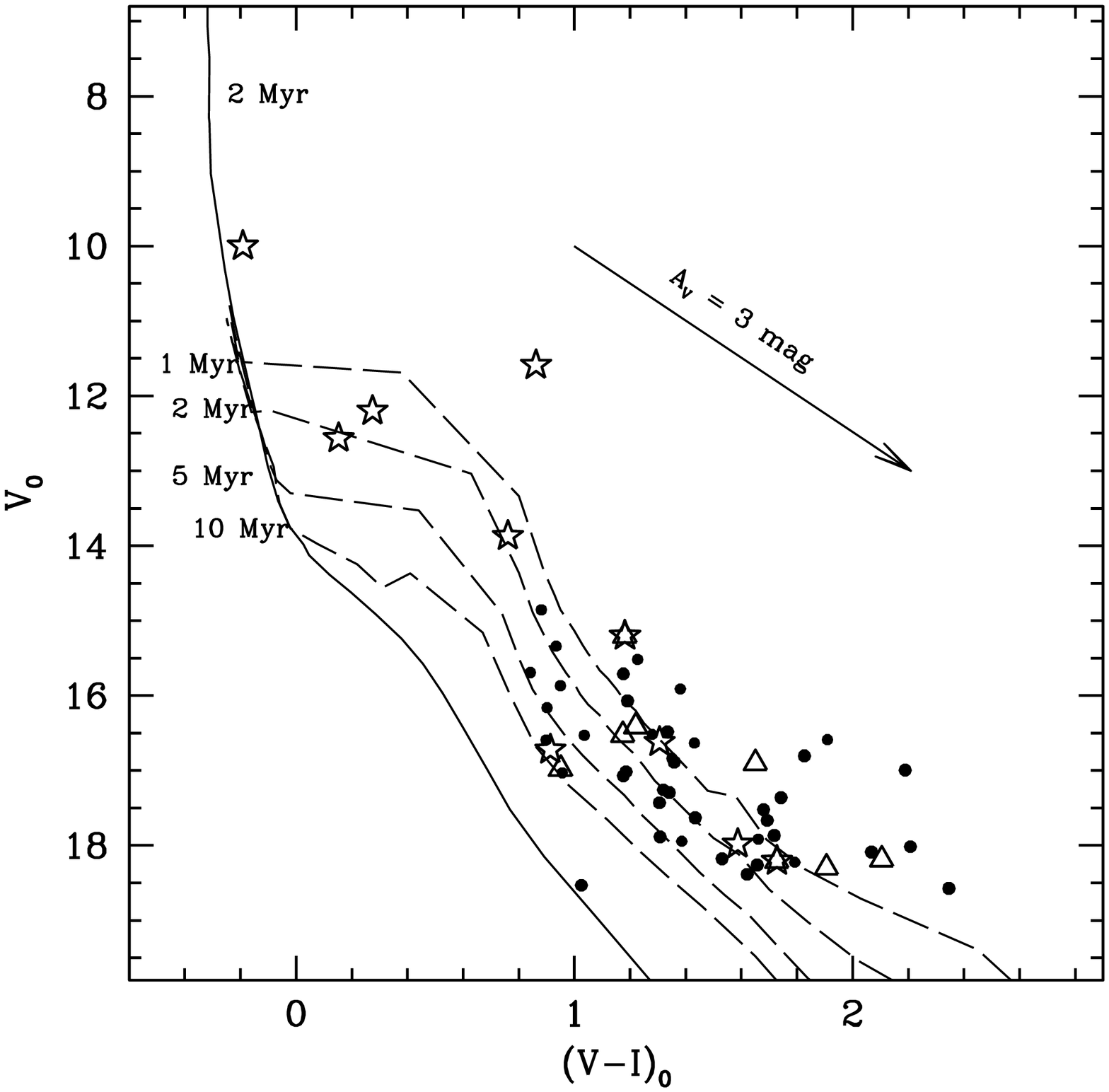}
\caption{Statistically cleaned $V_0/(V-I)_0$ CMD  for stars lying in the cluster region ({\it left panel}). 
Filled circles (Ages $\leq$ 5 Myr) are used to estimate the MF of the region.
{\it Right panel} shows the $V_0/(V-I)_0$ CMD for the $H\alpha$ emission stars (star symbols), NIR-excess stars (triangles)
and X-ray sources  in the `F' region of the NIR TCD (solid circles) (see \S 4.3 for details) collectively.
The ZAMS by Marigo et al. (2008) and the PMS isochrones
of 1,2,5,10 Myr along with evolutionary tracks for different mass by Siess et al. (2000) are also shown. All the curves are corrected for a distance of 2.81 kpc.}
\label{cleaned} 
\end{figure*}

The available distance estimates of NGC 281 in the literature varies from 2.0 kpc to
3.7 kpc (cf. Sato et al. 2008). 
For further analysis we adopt the VLBI trigonometric distance of the maser source of $2.81 \pm 0.24$ kpc (Sato et al. 2008).
Fig. \ref{cleaned} (left panel) shows statistically cleaned dereddened  $V_0/(V-I)_0$ CMD where stars having spectral type earlier than A0 were individually dereddened (cf. \S 4.2.1), whereas the mean reddening of the nearby region, estimated from the available individual reddening values in that region, was used for other stars. We have also plotted the ZAMS by Marigo et al. (2008) and the PMS isochrones by Siess et al. (2000) using the distance of  $2.81\pm0.24$ kpc.  The evolutionary tracks by Siess et al. (2000) for various masses have also been plotted which reveal that majority of the YSOs have  masses between 0.5-3.5 M$_\odot$. Fig. \ref{cleaned} (left panel) indicates an age spread for the PMS population. To check its reality, we plotted  $V_0/(V-I)_0$ CMD (assuming a mean $E(B-V)$=0.4 mag) for the H$\alpha$ emission stars, NIR-excess stars (probable CTTSs) and X-ray stars (probable WTTSs) (cf.  \S 4.3) in Fig. \ref{cleaned} (right panel). This also indicates an age spread of about 1-5 Myr for these probable PMS stars, supporting the reality of that in Fig. \ref{cleaned} (left panel). 

The age and mass of each YSO were estimated by comparing its location with the isochrones. 
Here we would like to point out that the estimation of the ages and masses of the PMS stars by
comparing their positions in the CMDs with the theoretical isochrones is prone to random as well 
as systematic errors (see e.g. Hillenbrand 2005, Hillenbrand 2008, Chauhan et al. 2009, 2011). 
Chauhan et al. (2009) and Barentsen at al. (2011) have studied the effect of random errors 
in the age estimation of PMS stars. Barentsen at al. (2011) found that uncertainty in the extinction 
estimation could play a significant role. In the case of NGC 281 region the variable extinction is small ($\sim 0.2$ mag), 
hence it should not contribute significantly in the errors. 
The effect of random error due to photometric error and reddening estimation in determination of ages and masses 
was estimated by propagating the random
errors to their observed estimation by assuming normal error distribution and using the Monte-Carlo simulations.
The estimated ages and their error are given in Table \ref{Tage}.
The systematic errors could be due to the use of different PMS evolutionary models and the error in distance estimation etc.
 Barentson et al. (2011) mentioned that the ages may be wrong by a factor of two due to the systematic errors in the model.
The presence of binaries may be the another source of error in the age determination.
Binarity will brighten the star, consequently the CMD will yield a lower age estimate.
In the case of an equal mass binary we expect an error of $\sim$ 50 - 60\% in the PMS
age estimation. However, it is difficult to estimate the influence
of binaries/variables on mean age estimation as the fraction of binaries/variables is not known. 
In the  study of TTSs in the  HII region IC 1396, Barentson et al. (2011) presumed that the number of binaries in their sample of TTSs could be very low as close binary lose their disc significantly faster than single stars (cf. Bouwman et al. 2006).

\begin{figure*}
\centering
\FigureFile(140mm,100mm){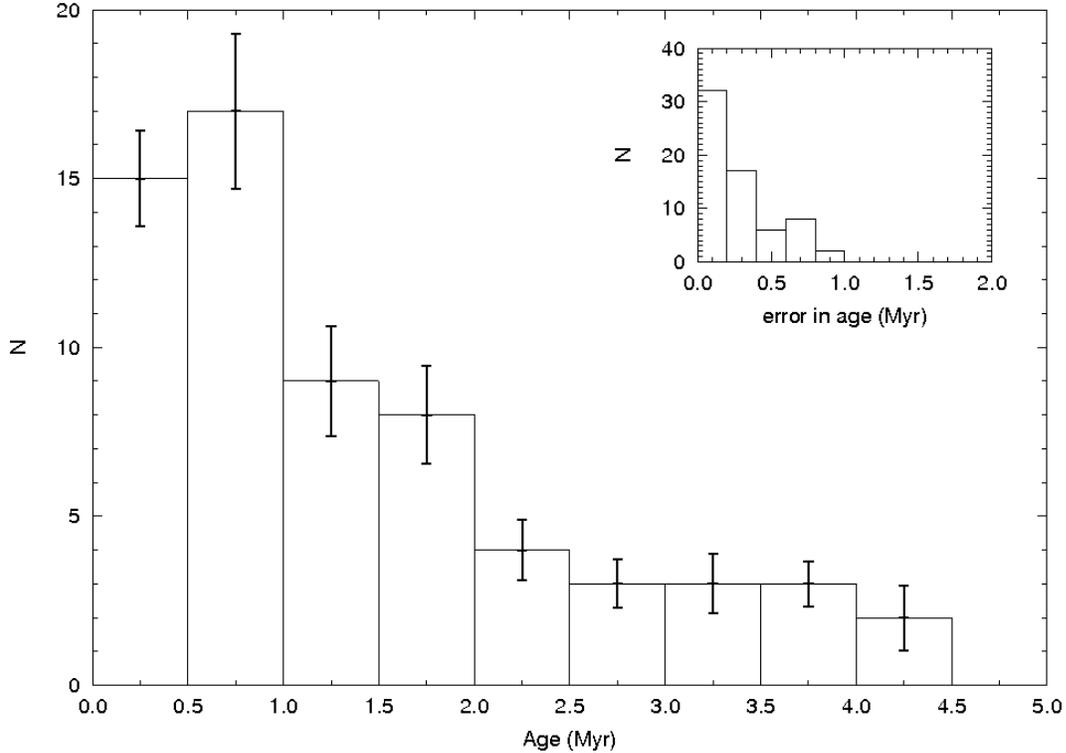}
\caption{\label{age}  Histogram showing the distribution of ages of the YSOs.  The inset shows the distribution of random errors.
 The error bar on Y-axis represent amount of scatter in each bin estimated on the basis of 
errors associated with age estimates as given in Table 9 and using Monte Carlo simulations.
}
\end{figure*}

The age distribution of YSOs shown in the Fig. \ref{age} indicates a significant scatter. The inset of Fig. \ref{age} 
shows the distribution of random errors. A comparison manifests that the age distribution of YSOs shows a significantly larger scatter
than that could occur due to random errors. 
Burningham et al. (2005) have investigated the effect of photometric variability in the apparent age spreads observed in the CMDs of OB associations. They found that the combination of binarity, photometric uncertainty and variability could not explain the observed age spread in the CMDs of OB associations.
If the effect of unresolved binaries is not significant,
we presume that the main reason of spread in the distribution of YSOs in Fig. \ref{cleaned} could be due to the spread in the ages of YSOs.  

The ages of young clusters are usually derived on the dereddened CMDs by comparing the earliest members to post-main-sequence evolutionary tracks if significant evolution has occurred and/or the low mass contracting population to the PMS isochrones. Since the most massive member of the cluster IC 1590 is an O6.5 MS star (Walborn 1973; Abt 1986; Guetter \& Turner 1997), its maximum age should be of the order of the MS life time of this star, i.e., $\sim$4.4 Myr (Meynet et al. 1994). Based also on the PMS stars, Guetter \& Turner (1997) derived a maximum age of $\sim$ 3.5 Myr for IC 1590.

We consider the points lying above the 5 Myr isochrone in Fig. \ref{cleaned} (left panel) as representing the statistics of the PMS stars in the cluster region. 
Here, we would like to remind the readers that the filled circles in Fig. \ref{cleaned} may {\it not} represent the actual members of the cluster. However, they should represent the statistics of PMS stars in the cluster region and this statistics is used to study its MF only.

\subsection {Initial Mass Function and K-band luminosity function}

The distribution of stellar masses that form in a star-formation event in a given volume of space is called initial mass function
(IMF). Young clusters are important  tools to study the IMF since they
are too young to lose a significant number of members either by dynamical or by stellar evolution.  
The MF is often expressed by a power law,
 $N (\log m) \propto m^{\Gamma}$ and  the slope of the MF is given as:

    $$ \Gamma = d \log N (\log m)/d \log m  $$

\noindent
 where $N  (\log m)$ is the number of stars per unit logarithmic mass interval. The classical value derived by Salpeter  (1955) is $\Gamma = -1.35$.

\begin{figure*}
\centering
\FigureFile(80mm,80mm){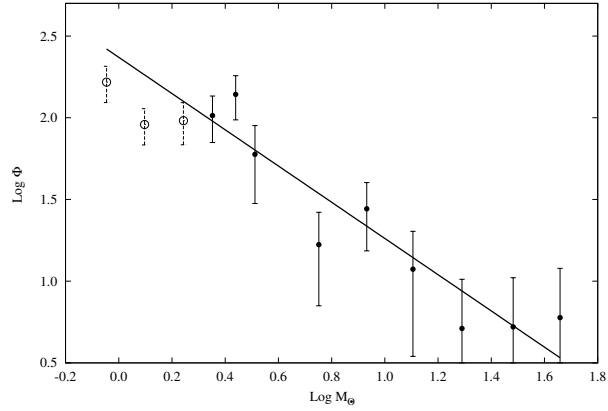}
\caption{ \label{mf} A plot of the mass function in the cluster. log $\phi$ represents log ($N$/dlog $m$). The error bars represent the $\pm\sqrt N$ errors. The solid line shows the least square fit to the mass range $2.0<M/M_\odot <54$ (shown by filled circles).}
\end{figure*}

With the help of the statistically cleaned CMD, shown in Fig. \ref{cleaned} (left panel),  
we can derive the MF using the theoretical evolutionary models. Since the age of the massive cluster members is thought to be $\sim$ 2 - 4 Myr, 
the stars having $V\lesssim13$ mag ($V_0 \lesssim 12$ mag; $M\gtrsim 4 M_\odot$) are considered to be still on the MS. For these stars, the LF was converted to a MF using the theoretical models by Marigo  al. (2008) (cf. Pandey et al. 2001, 2005).  The data for the three brightest stars, which were saturated in the present photometry, have been taken  from Guetter \& Turner (1997).
The MF for the PMS stars have been obtained by counting the number of stars in various mass bins (shown as evolutionary tracks)  having age $\leq$ 5 Myr in Fig. \ref{cleaned} (left panel). The resulting MF of the cluster is plotted in Fig. \ref{mf}. 
Since data incompleteness plays an important role in the estimation of the IMF, we restrict our analysis 
only to sources having $V<16.5$ mag. The present data have a completeness of $\sim$95\% at $V=$16.5 mag (cf. Table. \ref{completness}).
The slope ($\Gamma$) of the MF in the mass range $2 <M/M_\odot < 54$ comes out to be $-1.11\pm0.15$, 
which seems to be slightly shallower than the Salpeter (1955) value (-1.35). 
Using various combinations of the maximum expected errors in $E(B-V)$ and the distance, 
we found that the slope of the MF can vary in the range of $-1.08\pm0.15$ to $-1.16\pm0.15$. 
Keeping the errors in the estimation of $\Gamma$ in mind, it is difficult to decide whether the present
slope `$\Gamma$' is different from that of the Salpeter value. 
Guetter \& Turner (1997) have reported the slope of the MF ($\Gamma = -1.00\pm 0.21)$ for IC 1590 which, within error, 
is comparable to the value obtained in the present work.

\begin{figure*}
\centering
\FigureFile(80mm,80mm){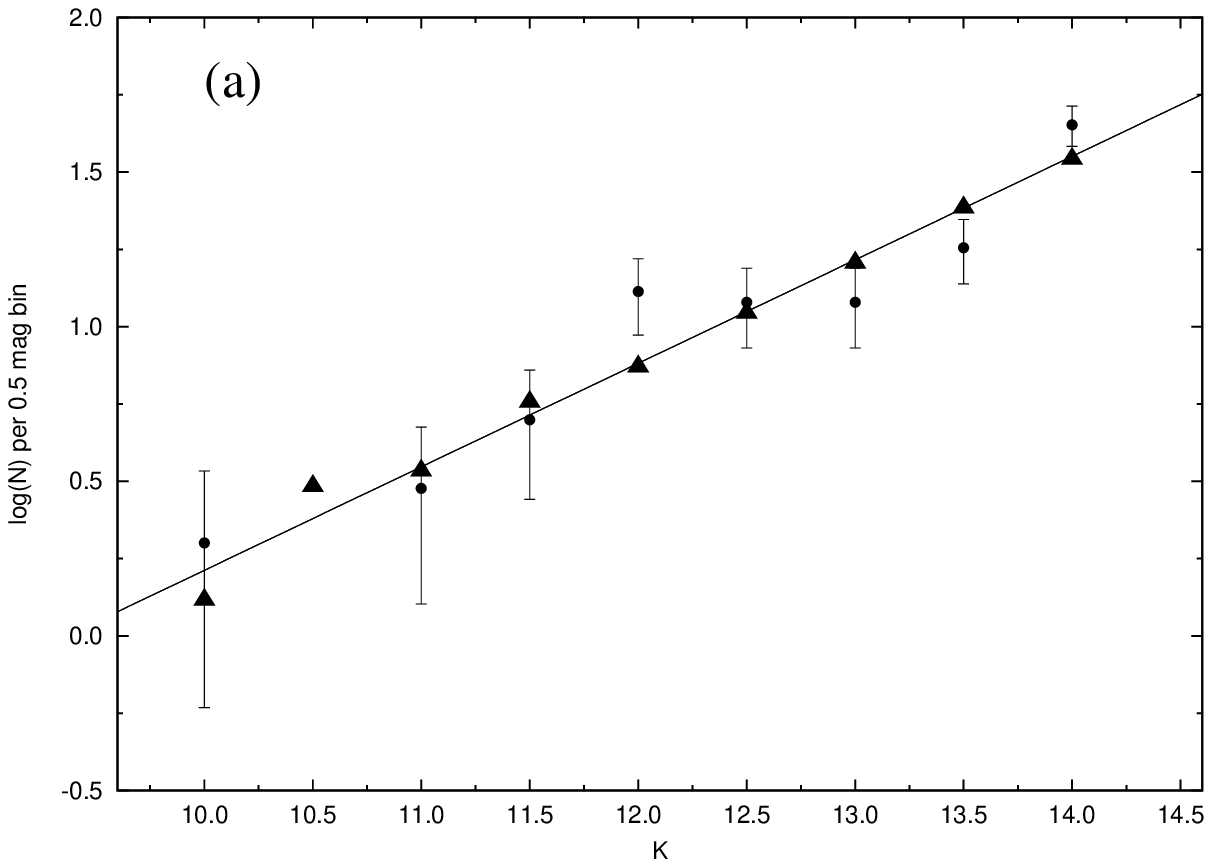}
\FigureFile(80mm,80mm){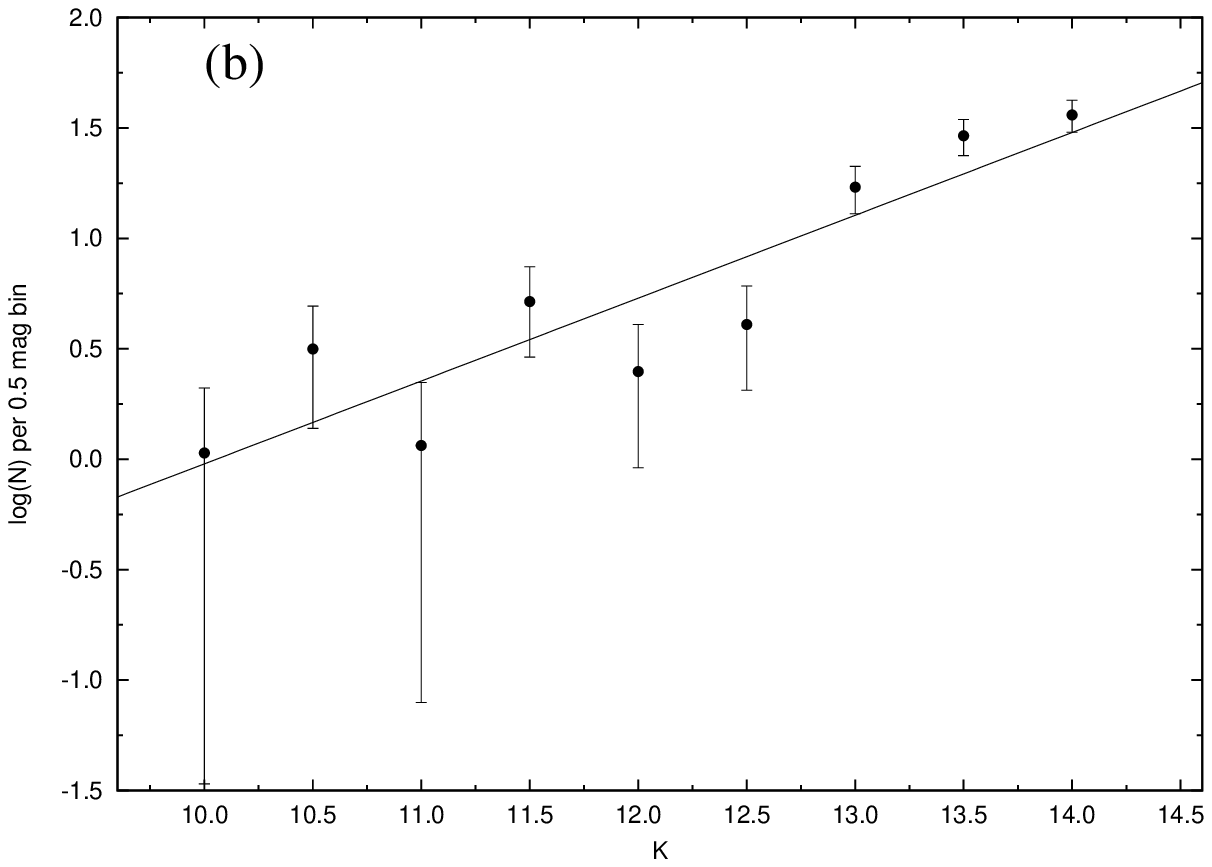}
\caption{\label{klf}(a) Comparison of the observed KLF in the reference field and the simulated
KLF from the star count modeling. The filled circles denote the observed $K$-band  star counts in the reference region, and the triangles represent the simulation from the Galactic model (see the text). The error bars represents the $\pm \sqrt N$ errors. The KLF slope ($\alpha$, see \S 4.5)
of the reference region (solid line) is $0.36\pm0.05$. The simulated model
is also gives a similar value of slope ($0.34\pm0.02$).
(b) The corrected KLF for the probable members in the cluster (see the text). The straight line is the
least-square fit to the data points in the magnitude range 10.5-14.25. The KLF slope ($\alpha$)
for the cluster region ($R_{cl}<5$ arcmin) is $0.37\pm0.07$. }
\end{figure*}

The K-band luminosity function (KLF) is a powerful tool to investigate the IMF 
of young embedded clusters; therefore during the last decade several studies focused on the
determination of the KLFs of young open clusters (e.g. Lada \& Lada 2003, Ojha et al. 2004b,
Sanchawala et al. 2007). In order to obtain the KLF of IC 1590, we again have to examine the effects of incompleteness and
field star contamination in our data. The completeness of the data is estimated using the
ADDSTAR routine of DAOPHOT as described in \S 2.1.1. To take into account the foreground/background
field star contamination, we used the Besan\c con Galactic model of stellar population
synthesis (Robin et al. 2003) and predicted the star counts in both the cluster region and in the direction of the reference field. 
We checked the validity of the simulated
model by comparing the model KLF with that of the reference field and found that the two KLFs match
rather well (Fig. \ref{klf}a). An advantage of using the model is that
we can separate the foregrounds ($d<2.8$ kpc) and the background ($d>2.8$ kpc) field stars.
The foreground extinction towards the cluster region is found to be $A_V \sim1.0$ mag.The model simulations with
$d<2.8$ kpc and $A_V$ = 1.0 give the foreground contamination, and that with $d>2.8$ kpc and $A_V$ = 1.7 mag the background population. 
We thus determined the fraction of the contaminating stars
(foreground+background) over the total model counts. This fraction was used to scale the nearby
reference region and subsequently the modified star counts of the reference region were subtracted
from the KLF of the cluster to obtain the final corrected KLF.
This KLF is expressed by the following power-law:

${{ \rm {d} N(K) } \over {\rm{d} K }} \propto 10^{\alpha K}$

\noindent
where ${ \rm {d} N(K) } \over {\rm{d} K }$ is the number of stars per 0.5 magnitude
bin and $\alpha$ is the slope of the power law. Fig. \ref{klf}b shows the KLF for the cluster region.
This indicates a slope of $\alpha = 0.37\pm0.07$ which is similar to the average 
slopes ($\alpha \sim 0.4$) for young clusters (Lada et al. 1991; Lada \& Lada 1995; Lada \& Lada 2003)
but higher than the values (0.27 - 0.31) obtained for Be 59 (Pandey et al. 2008)
and Stock 8 (Jose et al. 2008).

\section {Star formation scenario}

The star forming region NGC 281 has been attracting attention of the star forming community. It contains a cluster IC 1590 with a Trapezium like system of O-type stars at the center.
The ionized hydrogen seems to be associated with two CO molecular
clumps (east and west) which have been mapped in $^{12}$CO and $^{13}$CO by Elmegreen \& Lada (1978), 
Leisawitz, Bash \& Thaddeus (1989), Henning et al. (1994), Megeath \& Wilson (1997) and Lee \& Jung (2003). The western CO clump called NGC 281 West is somewhat more massive and compact than the 
elongated eastern clump (NGC 281 East) (Lee \& Jung 2003).

The ionizing source HD 5005 lies to the northeast/northwest of NGC 281 West/NGC 281 East.
The differential extinction towards the central cluster is $\sim$0.2 mag (cf. \S 4.2.1) 
indicating that the central cluster contains only gas and dust of low-density.  A similar trend has been noticed in many
clusters associated with HII regions (e.g. 30 Dor, Brandl et al. 1996; NGC 3603, Pandey et al. 2000 and  NGC 1893, Sharma et al. 2007). 
A reasonable explanation for this lack of a dense medium in the central region may be 
the effects of the massive star(s) at the cluster center. 

As indicated in previous studies (cf. Lee \& Jung 2003 and references therein), the western molecular clump is 
interacting with the ionized gas. On the basis of kinetic evidence, Elmegreen \& Moran (1979) suggested the
passage of a shock through NGC 281 West. An H$_2$O maser was found to be coincident with the peak of the cloud 
indicating the ongoing star formation (Elmegreen \& Lada 1978). Carpenter et al. (1993) and Megeath (1994) detected a cluster of low mass stars associated with NGC 281 West. Elmegreen \& Lada (1978) suggested that this region is a site of triggered star formation through the ``collect and collapse'' mode. However, Megeath \& Wilson (1997) claimed that numerical models of imploding spherical clumps can approximately reproduce the kinematic features observed in NGC 281 West, and suggested that ``radiation-driven implosion" (RDI) is a more plausible and attractive model. 
Between NGC 281 East and the central cluster a few bright-rimmed clouds (BRCs) or cometary globules and 
an IRAS source are located. Some of the H$\alpha$ stars can be seen around the tip of the BRCs just as 
observed in several well-known BRCs (see Ogura et al. 2002). 
Both the CO clumps show the presence of YSOs (IR-excess and H$\alpha$ stars) around their respective centers. 

Whatever the star formation scenario, the ionization/shock fronts caused by high-mass stars of the first generation appears to have initiated the formation of a new generation of stars at the edge of the molecular clumps. The distribution of YSOs and morphological details of the environment around the cluster can be used to infer the star formation history of the NGC 281 region in detail. To know the distribution of YSOs in the region, we have included all the detected YSOs without considering their photometric errors to improve the sample. 

\begin{figure*}
\centering
\hbox{
\FigureFile(80mm,80mm){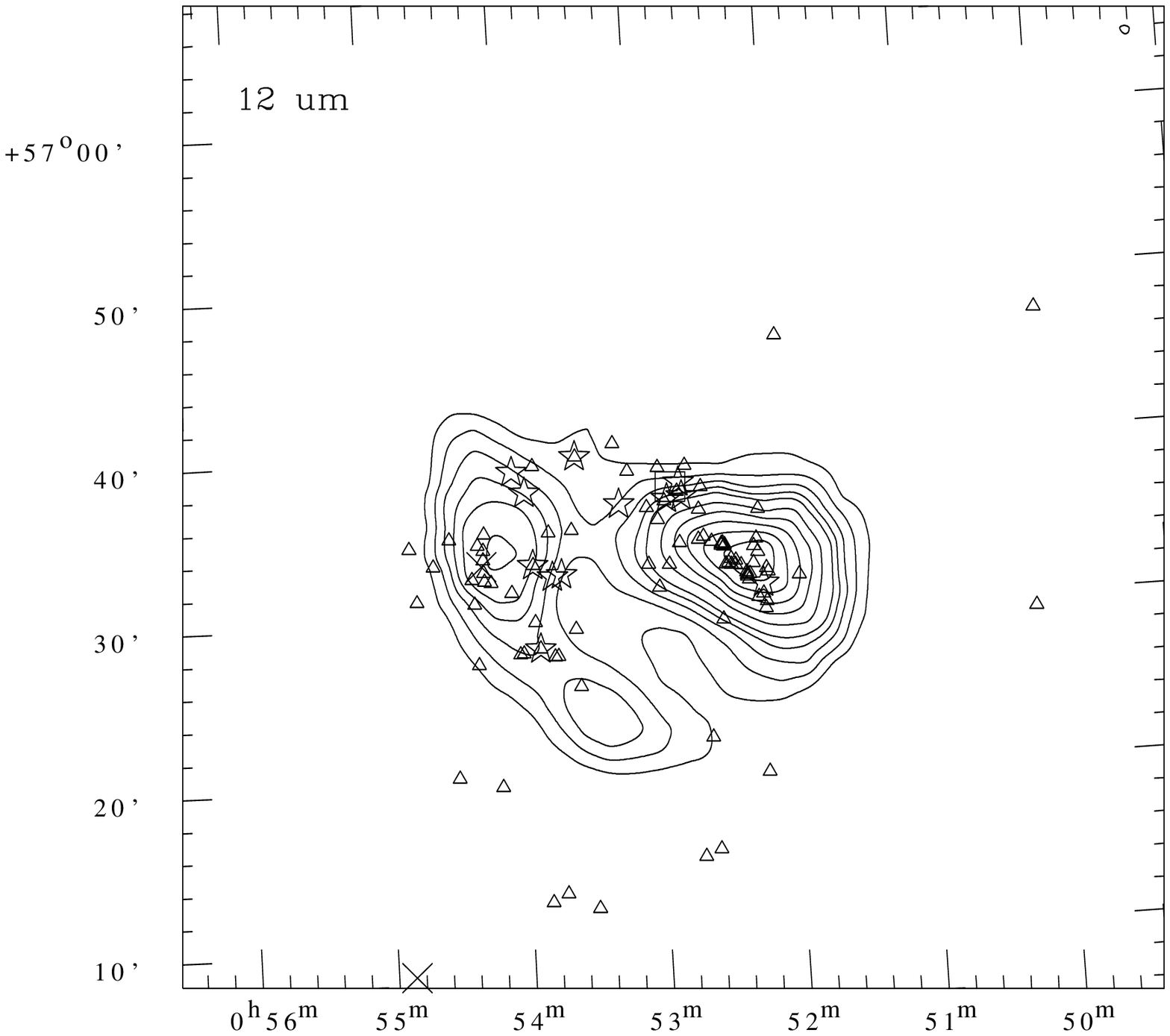}
\FigureFile(80mm,80mm){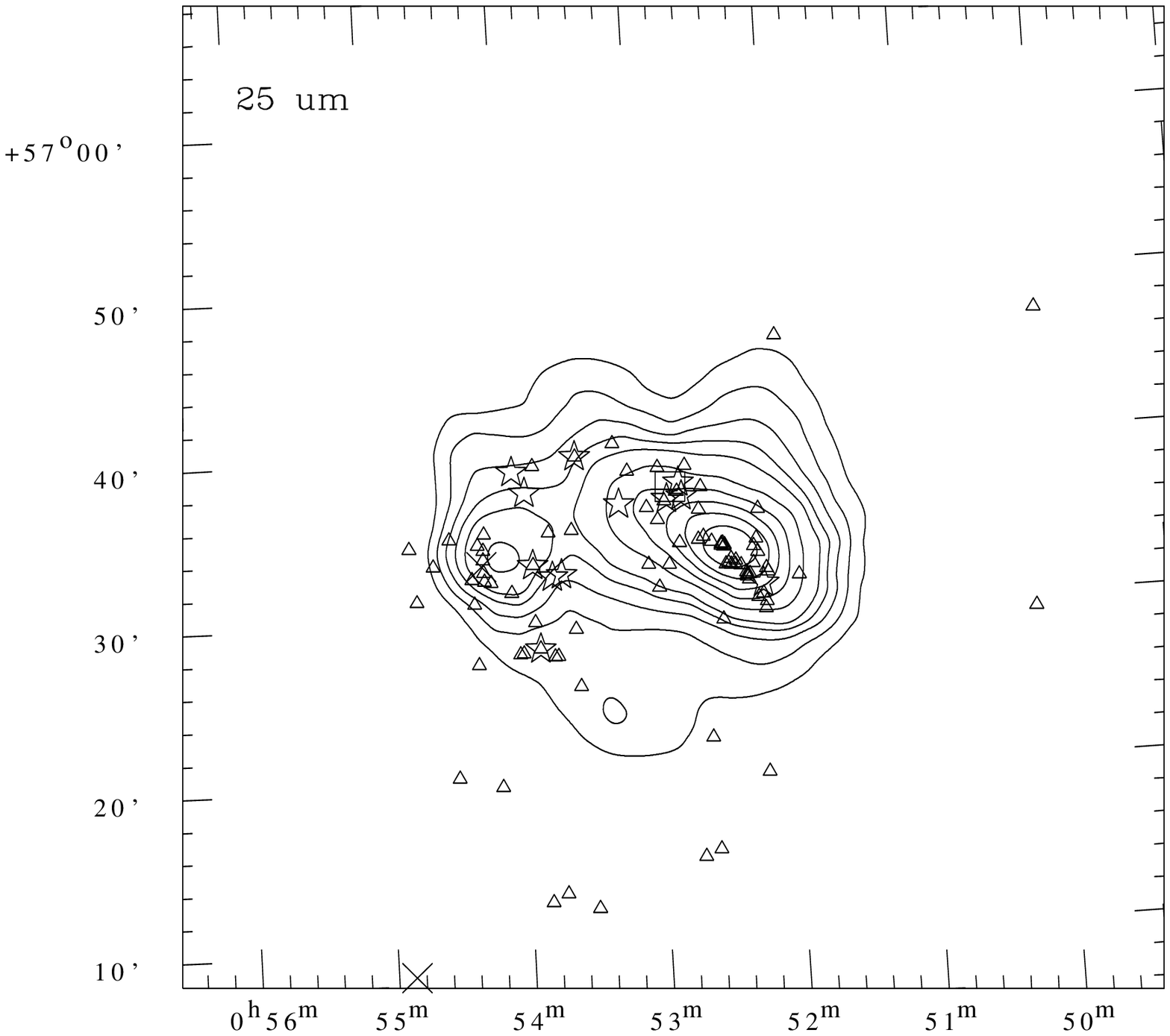}
}
\hbox{
\FigureFile(80mm,80mm){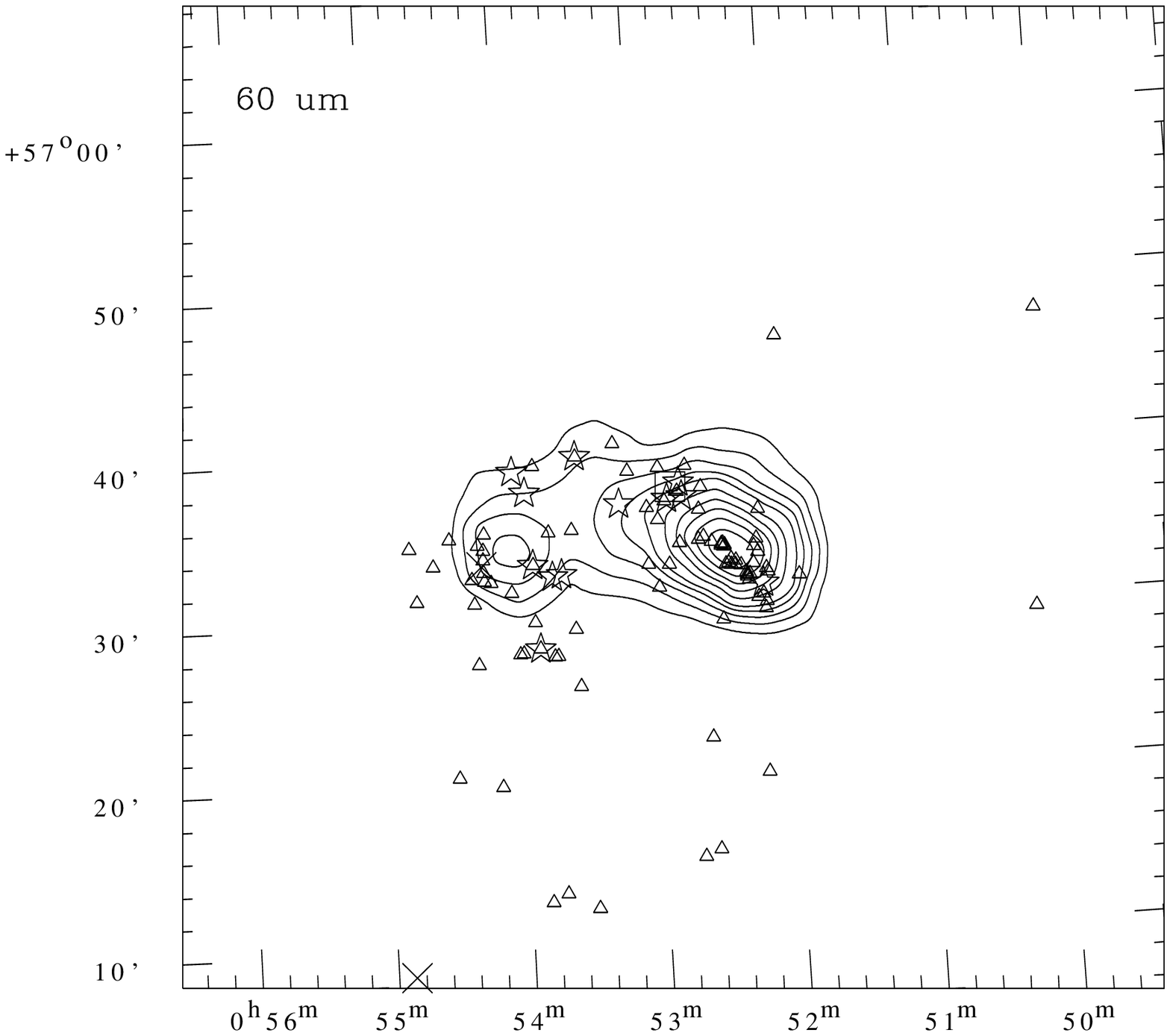}
\FigureFile(80mm,80mm){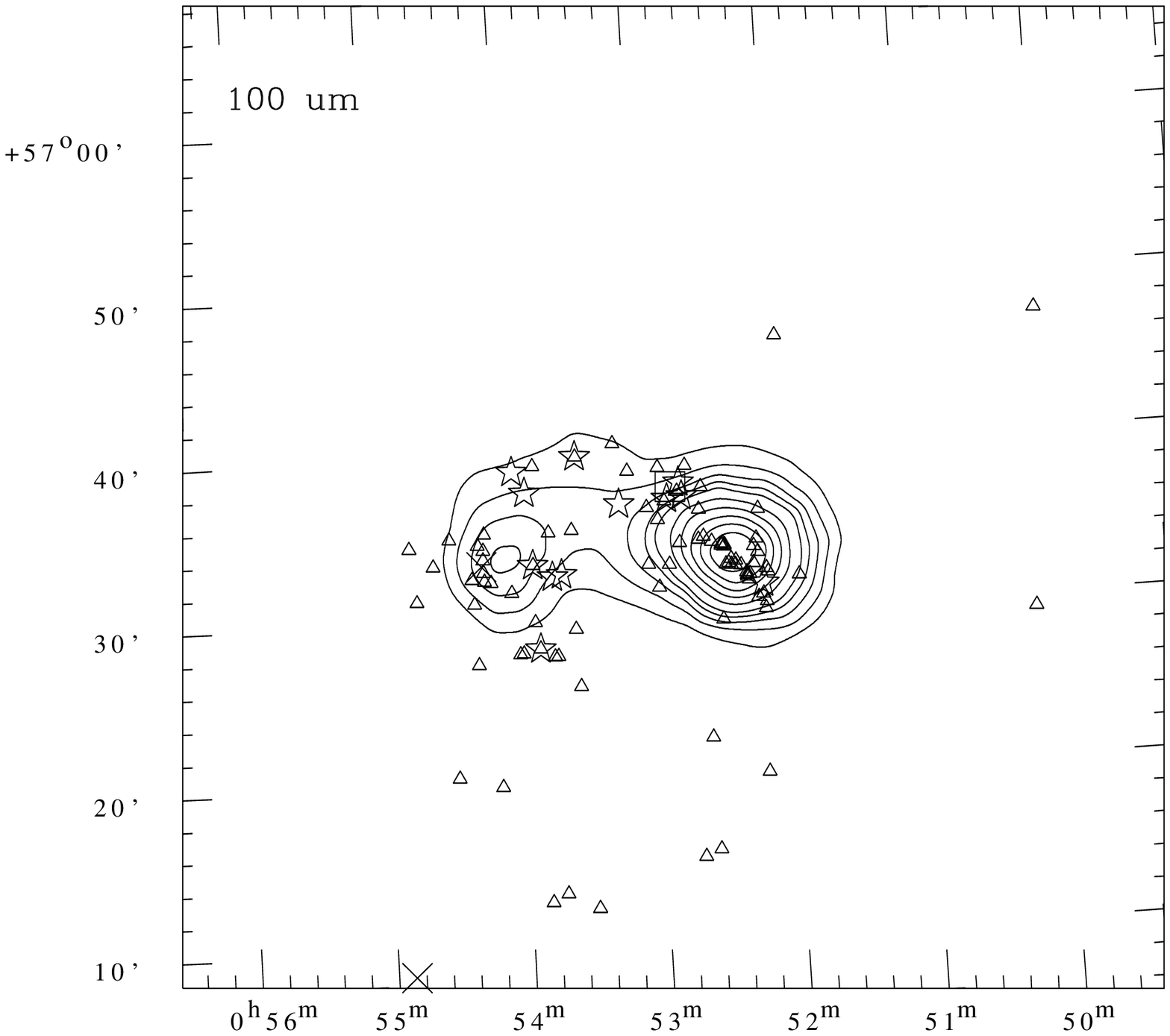}
}
\caption{\label{hires} The IRAS intensity maps for the cluster region in $12~\mu m~ (top~ left)$, $25~ \mu m~ (top~ right)$,
$60~ \mu m~ (bottom~ left)$ and $100~ \mu m~ (bottom~ right)$.
The contours are at 20, 25, 30, 35, 40, 50, 60, 70, 80 and 90 \% of the peak value of 20 MJy/ster, 48 MJy/ster,
412 MJy/ster and 907 MJy/ster in 12, 25, 60 and 100 $\mu m$ respectively. 
The locations of IR-excess stars (probable CTTSs, triangles), $H\alpha$ emission stars (star symbols), 
IRAS point sources (crosses), and the O-type star (open square) are also shown in the  image.
The abscissa and the ordinates are in the J2000. }
\end{figure*}

IRAS maps can be used to study the distribution of dust and unidentified infrared band (UIB) carriers. Fig. \ref{hires} shows the IRAS intensity maps for the NGC 281 region at 12 $\mu$m (top left), 25 $\mu$m (top right), 60 $\mu$m (bottom left) and 100 $\mu$m (bottom right). The global features of these maps are quite similar to that of the molecular gas. All the contours show peaks around both the eastern and western clumps indicating the coexistence of  warm dust (IRAS 12 and 25 $\mu$m), cold dust (IRAS 60 and 100 $\mu$m) and molecular gas. NGC 281 East lacks radio continuum emission and its extended distribution of the 12 $\mu$m emission towards the south resembles that of the molecular material. 
As pointed out by Leisawitz, Bash \& Thaddeus (1989), in the case of eastern clump, the peak of ionized gas (as seen from the ionized source) is followed by the peaks of IR emission and CO emission from the molecular cloud, respectively.
The coincidence of the peak of  the IRAS and CO emission, the location of the IRAS point source and the distribution of YSOs indicate ongoing star formation activity in these clumps.

\begin{figure*}
\centering
\FigureFile(160mm,160mm){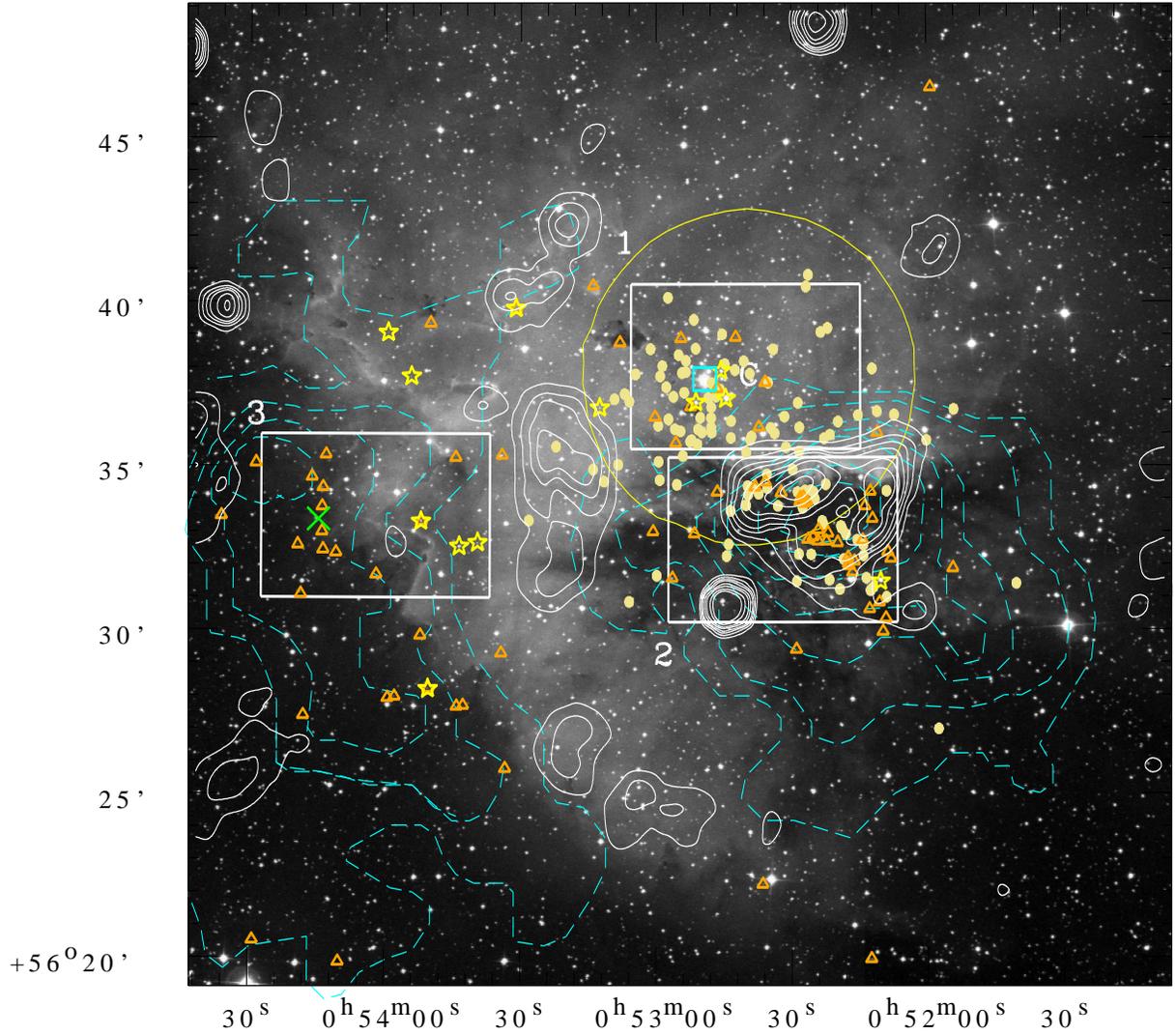}
\caption{\label{spa30} The locations of IR-excess stars (probable CTTSs, 
triangles), $H\alpha$ emission stars (star symbols), IRAS point sources (crosses), the O-type star (open square),
and WTTS (filled circles) are overlaid on the DSS-2 $R$ band $30 \times 30$  arcmin$^2$  image. 
The cluster region is represented by the circle having the center ``C''.
The CO contours  taken from Henning et al. (1994) are shown by the {\it blue dashed lines} and the
NVSS ($1.4~GHz$) radio continuum contours by the {\it white lines}. The radio contours are 5, 10, 15, 20, 30, 
40, 60, 80\% of the peak value of 0.04 Jy/Beam.  Three subregions are also marked by the boxes.  
The abscissa and the ordinate are in J2000.  }
\end{figure*}

\begin{figure*}
\centering
\hbox
{
\FigureFile(80mm,80mm){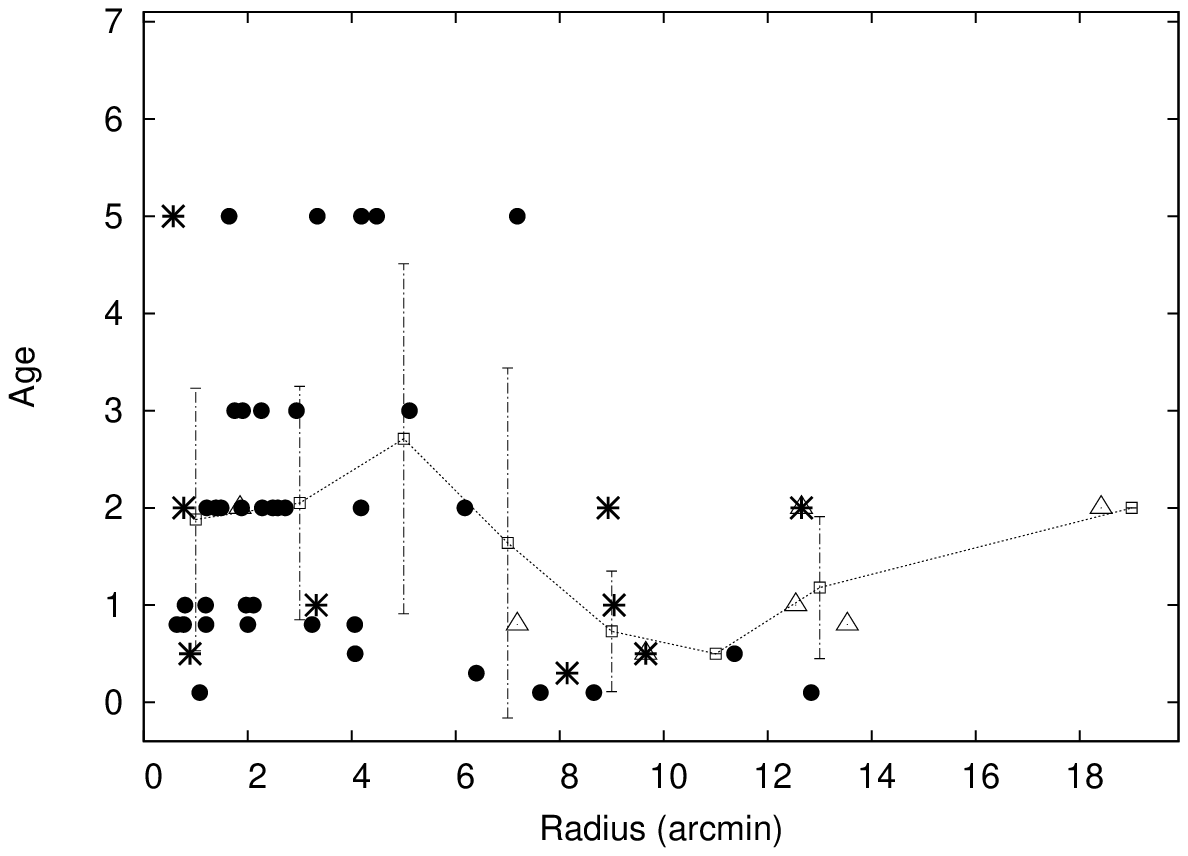}
\FigureFile(80mm,80mm){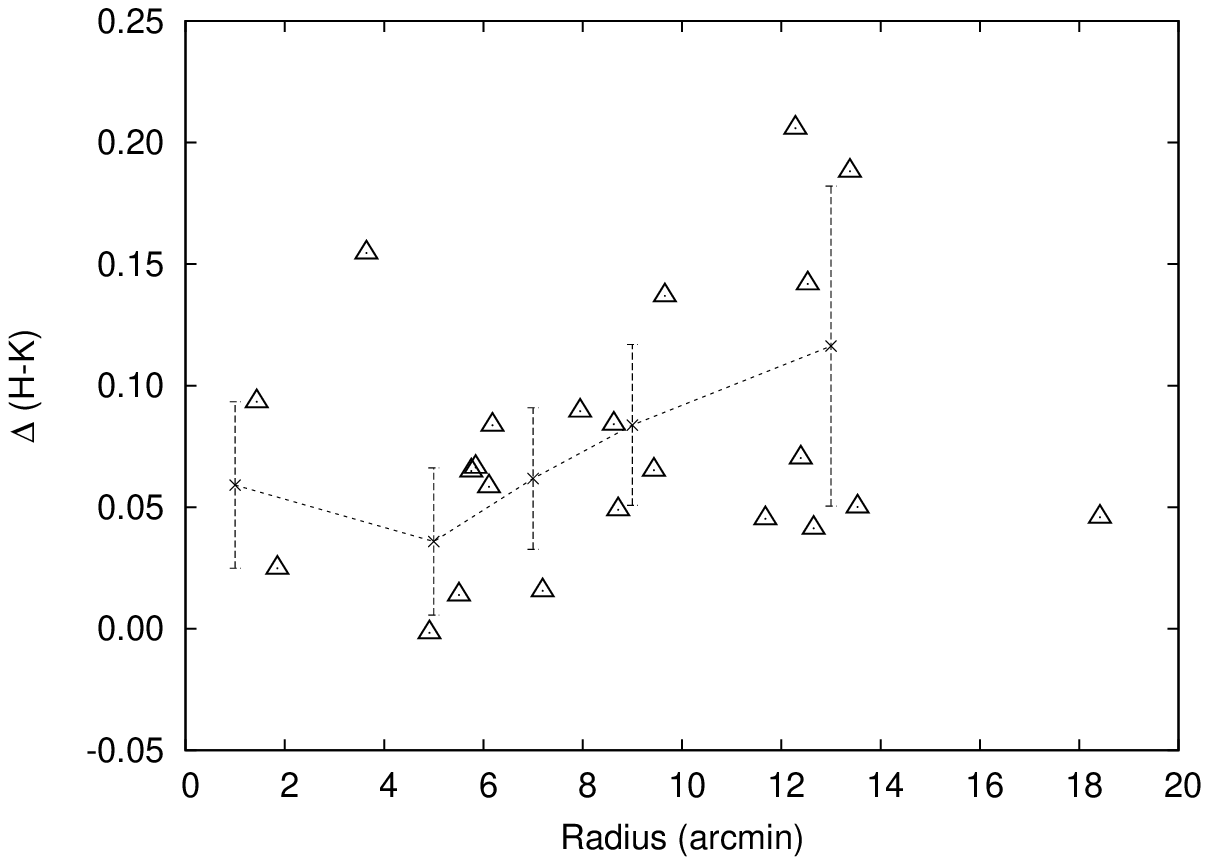}
}
\caption{\label{age-hk} Variation of the age  and  NIR-excess [$\Delta (H-K)$] of the YSOs as a function of radial distance from the O star HD5005.
The symbols are same as in Fig. \ref{nir-yso}.
The dotted line represents the $Mean \pm \sigma$ of the distribution.
}
\end{figure*}

Fig. \ref{spa30} shows the map of the $^{12}$CO emission taken from Henning et al. (1994) and the 1.4 GHz radio emission from NVSS along with the spatial distribution of all the detected YSOs overlaid on the DSS-2 $R$ band image. The center of the cluster IC 1590 is marked by `C' and the location of the ionization source is shown by a square.  A well aligned distribution of the detected YSOs from the vicinity of the ionization source to the direction of NGC 281 West can be noticed. This spatial distribution of the YSOs resembles with that in the case of NGC 1893 and BRC 14, where a similar distribution of NIR-excess stars can be noticed from the ionization source to the direction of the cometary globules/BRC (see figure 22 of Sharma et al. 2007 and figure A3 of Chauhan et al. 2009). These alignments in NGC 1893 and BRC 14 were attributed to triggered star formation due to a series of RDI process. In both the cases YSOs located away from the ionization sources are found to be younger. Sicilia-Aguilar et al. (2004) have also shown that in the case of the Tr 37/IC 1396 globule region, CTTSs are aligned from the iosizing source  towards the direction of the globule, and that most of the younger ($\sim$1 Myr) members appear to lie near or within the globule. They mentioned that it can be indicative of triggered star formation. Fig. \ref{age-hk} (left panel) shows the age distribution of the YSOs as a function of the radial distance from HD 5005, the ionization source of NGC 281,
which shows that the YSOs in the cluster region ($r \lesssim 8^\prime$) have ages $\lesssim$ 5 Myr whereas those lying outside the cluster
region are relatively younger and have ages $\lesssim$ 2 Myr. 
Fig. \ref{age-hk} (right panel) shows the radial variation of the NIR-excess $\Delta (H-K)$, 
which is defined as the horizontal displacement from the middle reddening vector at the 
boundary of  the `F' and `T' regions (see Fig. \ref{nir-yso}).
To quantify the radial variation of the age and NIR-excess, we used only those stars which have the error $\le$ 0.1 mag.
The distribution of $\Delta (H - K)$ also suggests that the sources lying outside the boundary of the cluster ($r\sim8^\prime$) 
have relatively larger NIR excess in comparison to those located within the cluster region. 
However we admit that the above statements are not conclusive in view of the fact that the scatters are large and the differences are subtle.
A deeper optical, NIR and MIR observations are needed to have conclusive star formation scenario in the region. 
A similar trend has been reported in the case of a few BRCs by Chauhan et al. (2009)  as well as in a recent study on the IC 1396 region by Barentsen et al. (2011).

The near-IR excess in the case of CTTSs suggest the presence of dusty optically thick discs (Haisch
et al. 2001a, Sicilia-Aguilar et al. 2006). Sicilia-Aguilar et al. (2006) found similarities 
between the decrease in IR excesses and the decrease in accretion
rates, and concluded that gas evolution seems to occur somehow parallel to the evolution of the dust
grains and the structure of the disc. In a recent study, Sicilia-Aguilar et al. (2009) further confirmed
that IR-excess decreases with age. Hence, relatively large NIR excess outside the cluster region (cf. Fig. \ref{age-hk}, right panel) 
could be indicative of relatively younger population outside the cluster region.

\begin{figure*}
\centering
\FigureFile(160mm,60mm){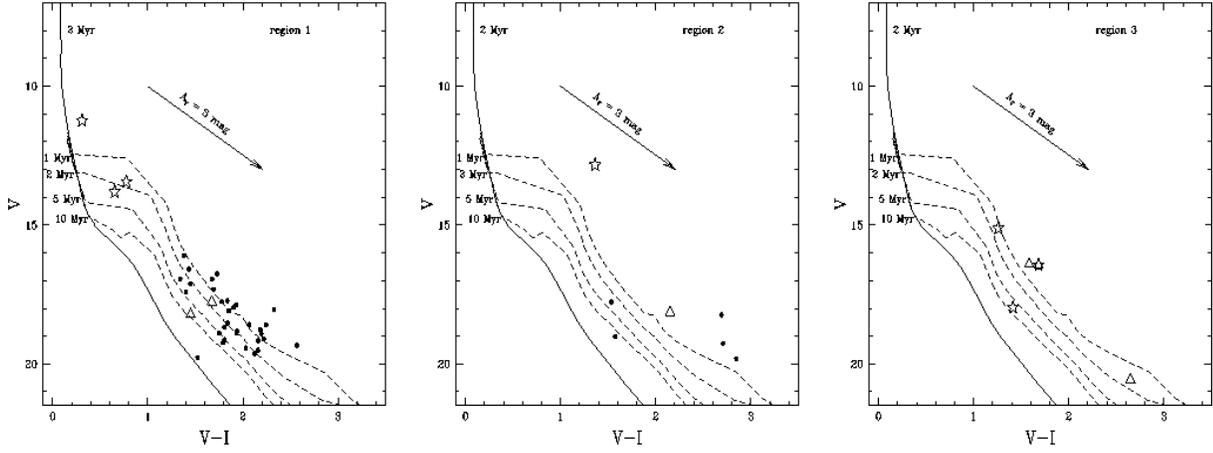}
\caption{\label{region-cmd}
$V/(V-I)$ CMDs for the probable YSOs detected in three subregions. The symbols are the same as in Fig. \ref{nir-yso}.
The isochrone for 2 Myr (continuous curve) by Marigo et al. (2008) and the PMS isochrones
(dashed curves) for ages 1,2,5,10 Myr by Siess et al. (2000) are also shown. 
All the isochrones are corrected for the reddening [$E(B-V)_{min}$ = 0.32 mag] and the distance of 2.81 kpc.
}

\end{figure*}

\begin{figure*}
\centering
\FigureFile(160mm,60mm){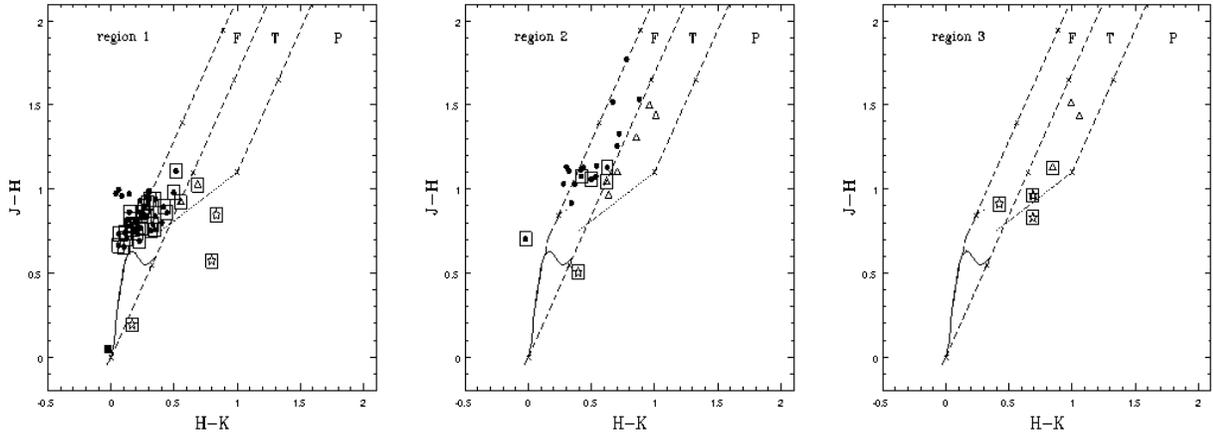}
\caption{\label{region-ccd} NIR TCDs of the YSO candidates in the three subregions.
The sequences for dwarfs (solid curve) and giants (thick dashed curve)
are taken from Bessell \& Brett (1988). The dotted line represents the loci of unreddened T Tauri stars
(Meyer et al. 1997). Dashed straight lines represent the reddening vectors (see the text).
The crosses on the dashed lines are separated by $A_V$ = 5 mag. 
Open squares show the data having optical counterparts used in Fig. \ref{region-cmd}.
The other symbols are as same as in Fig. \ref{nir-yso}.
}
\end{figure*}

To study the evolutionary stages of the NGC 281 region, we have divided it into three sub-regions as shown in Fig. \ref{spa30}, namely Region 1 (IC 1590), Region 2 (NGC 281 West) and Region 3 (NGC 281 East). Figs \ref{region-cmd} and \ref{region-ccd} show the $V/(V - I)$ CMDs and NIR TCD, respectively, for the three regions. Here also we have used only sources of good quality, i.e., sources having error less than 0.1 mag. X-ray data are not available for Region 3. The CMDs indicate that the age of the YSOs in Region 1 ranges between $\lesssim$ 1 - 5 Myr, whereas a majority of the YSOs in Region 2 have ages $<$ 1 Myr.  The YSOs associated with Region 3 indicate an age of $\sim$ 1-2 Myr for the region. The NIR TCDs (Fig. \ref{region-ccd}) indicate that the extinction in Regions 2 and 3 is relatively higher in comparison to Region 1.

A comparison of the statistics of the YSOs in these three regions can also give a clue about the evolutionary stages of the regions. Table \ref{statT} shows the statistics of the probable CTTSs associated with the three sub-regions. We have given statistics for the sources having error $\le$ 0.1 mag as well as for  all the detected YSOs without considering their errors. The total number of stars detected in the 2MASS Catalogue and the number of field stars expected in each region, estimated from the nearby reference region (cf. \S 4.1.2), are also given in Table \ref{statT}. 
The percentage of H$\alpha$ stars and NIR excess stars in each region is estimated after substracting the contribution of expected field stars
(89 stars. column 2 of Table \ref{statT}) from the total stars (also given in column 2 of Table \ref{statT}) in the subregions.
The fraction of detected CTTSs  (i.e. NIR-excess stars and  H$\alpha$ stars), for both the samples, is significantly higher in Regions 2 and 3  as compared to the cluster region (Region 1). This is further supported by a comparison of the CTTSs fractions, {\it f$_{CTTS}$} = {\it N$_{CTTS}$}/{\it (N$_{CTTS}$} + {\it N$_{WTTS}$)} in Region 1 (5-12 \%) 
and Region 2 (25-45 \%) (cf. Table \ref{ctts}). 
The sources flagged as 1 and 2 in column 14 of Table \ref{sltt} are classified as CTTSs and WTTSs respectively. 
The WTTSs are identified using the NIR TCD and X-ray data (cf. \S 4.3.1). The X-ray data is not available for the Region 3. 
Here it is important to mention that as mentioned in \S 4.3.1 some of the CTTSs having less or negligible NIR excess might have been classified as WTTSs. If this is true, the {\it f$_{CTTS}$} will have further higher value.
Haisch et al. (2001b) have found the disc evolution fraction in the sense that the fraction of stars 
having a disc decreases with age. Armitage et al. (2003) have also found that,  in Taurus-Auriga T-association, {\it f$_{CTTS}$}, decreases with stellar age. A comparison of the disc fractions
 of Regions 1 and 2 with those given by Haisch et al. (2001b) and Armitage et al. (2003) suggests ages
of $\sim$ 4-5 Myr and  $\sim$ 1-3 Myr for Regions 1 and 2 respectively, which is in fair agreement with the results obtained from the CMDs. 
Thus the above discussions suggest the propagation of (triggered) star formation in Regions 2 and 3.

\begin{figure*}
\centering
\FigureFile(140mm,160mm){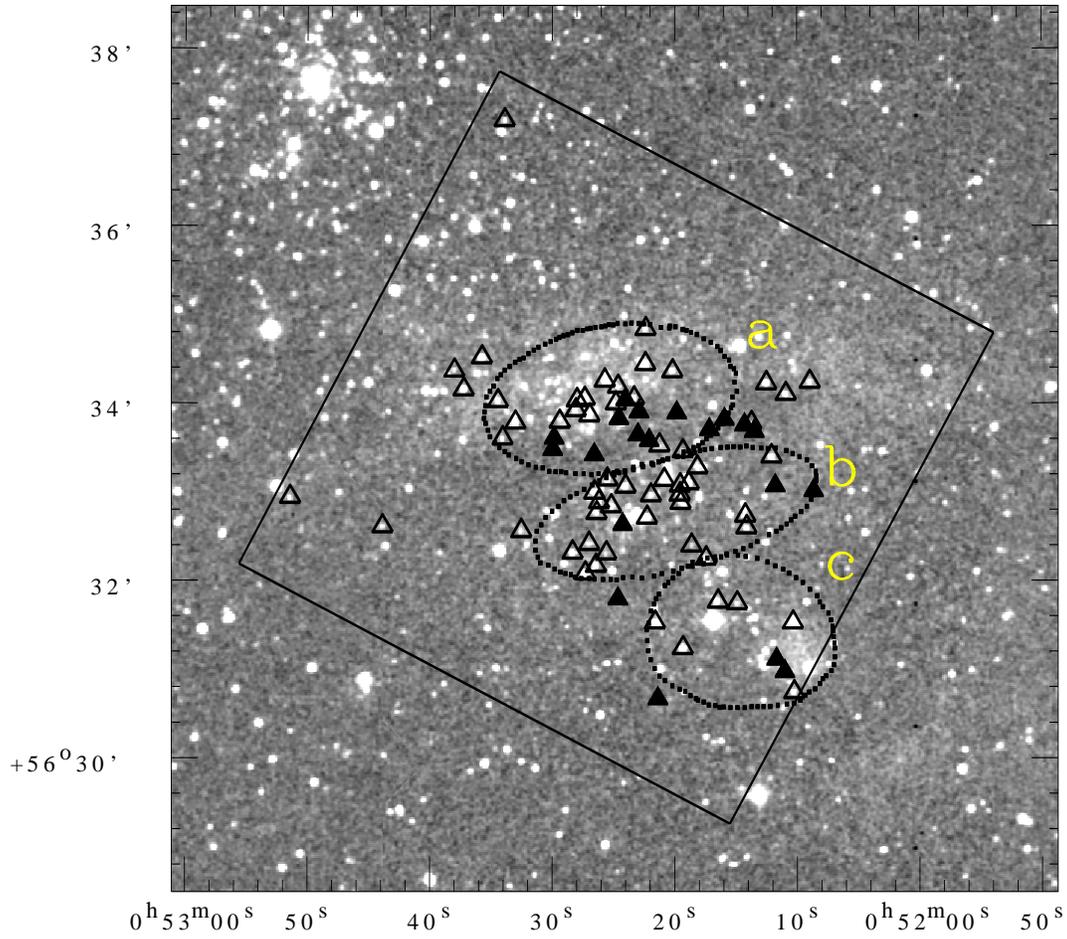}
\caption{\label{2m-spit} Spatial distribution of the Class 0/I (filled triangles) and Class II (open triangles) 
sources detected by {\it Spitzer} IRAC overlaid on the 2MASS image of the  NGC 281  region.
The square box represents the area observed by {\it Spitzer} IRAC. 
The dashed curves indicate the boundaries of the three sub-clusters named as a, b and c (see \S5 for details). }
\end{figure*}

Megeath \& Wilson (1997) pointed out the presence of two sub-clusters (northern and southern) in NGC 281 West. 
They concluded that the northern and southern sub-clusters are indeed physically separate sub-clusters 
resulting from distinct star formation events. Based on the location of the sub-clusters they concluded that 
the northern sub-cluster appears to be associated with the NW and NE clumps, whereas the southern 
sub-cluster seems to be associated with S clump.
The spatial distribution of the Class 0/I and Class II sources detected by {\it Spitzer} observations overlaid on the 2MASS image is shown in Fig. \ref{2m-spit}. Certainly the YSOs are found to make two sub-clusters. However, a careful look of this figure manifests a third sub-clustering also. The isodensity contours shown in Fig. \ref{iso} also suggest the presence of three sub-clusters. We have visually marked the boundaries of these sub-clusters, designated as `a', `b' and `c'.
 Some of the detected YSOs are located outside the boundaries of these sub-clusters. Megeath  \& Wilson (1997) also 
found that northern sub-cluster has higher extinction as compared to the southern sub-cluster. 
A comparison of the morphology of the molecular cloud as observed in C$^{18}$O (2-1) 
(Megeath \& Wilson 1997) and  the distribution of YSOs in the `a' sub-cluster region indicates that `a' sub-cluster 
seems to be associated with the NW and NE clumps as referred by Megeath \& Wilson (1997), whereas `c' sub-cluster seems to be associated with the clump `S'.
 Megeath \& Wilson (1997) found that the clump `S' (-34 to -32 kms$^{-1}$) is kinematically and spatially distinct 
from the NW and NE clumps (-32 to -29 kms$^{-1}$). The velocity distribution suggests that clump `S' (i.e. sub-cluster `c') 
is relatively near to the observer in comparison to the NW clump `a'. 
If we assume that the sub-clusters are not associated with each other, the spatial distribution of the YSOs in the sub-cluster `a' reveals
that the Class II sources are relativley near to the ionizing source in comparison to the Class 0/I sources. 
We do not find this trend in the sub-cluster `b'. The distribution of YSOs in sub-cluster `c' also show a similar trend however the statistics is poor.
The distribution of the YSOs detected by using the IRAC data in the case of a few BRCs also shows that the Class 0/I sources are found to be located more away from the ionizing sources as compared to Class II sources.  This seems to indicate the propagation of star formation in the NGC 281 West region.

\section{PMS sources: MIR and X-ray observations}

\begin{figure*}
\centering
\FigureFile(90mm,90mm){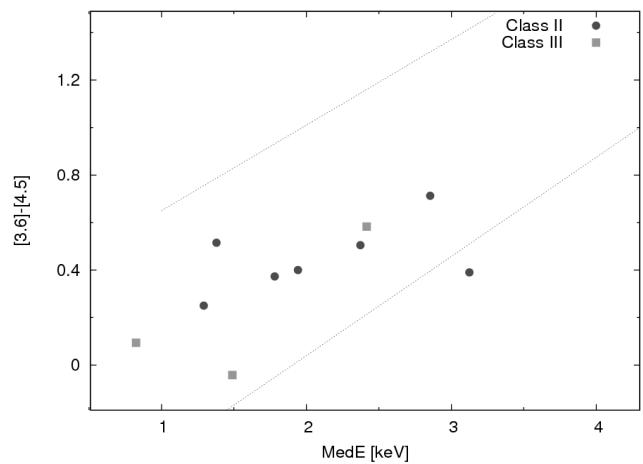}
\caption{\label{Med}   {\it Spitzer} [3.6]-[4.5] MIR colour vs. {\it Chandra} X-ray source median energy. 
Filled circles and squares represent Class II and Class III sources, respectively.. The dashed lines indicate 
the boundaries of the distribution of similar objects in the IC 1396N region (Getman et al. 2007). 
 }
\end{figure*}

As discussed in \S 4.3.2 that none of the Class 0/I sources show X-ray emission. The high value of $A_V$ in the region
could be a possible reason for non-detection of X-ray in the Class 0/I sources.
X-ray median energy (MedE) is a reliable indicator of absorbing column density through the empirical relation,   log $N_{\rm H}$ = 21.22 + 0.44 (MedE) cm$^{-2}$ (Feigelson et al. 2005). Getman et al. (2007) have demonstrated a relationship between MedE and the MIR colour [3.6]-[4.5]  for obscured PMS stars. Fig. \ref{Med} shows a MedE vs. [3.6]-[4.5] diagram. The dashed lines represent the boundaries of the distribution  obtained in  IC 1396 N by Getman et al. (2007, see their figure 8). Using the {\it Spitzer} photometry and assuming that the relationship between MedE and MIR colours given by
Getman et al. (2007) is valid for the NGC 281 region also, we estimated the MedE for the Class II/III sources detected in the present study as $\lesssim$ 4 keV. Using the above relation, the absorbing column densities towards these sources have been estimated as  $N_H$ $\gtrsim$ 10$^{23}$ cm$^{-2}$, which yields $A_V$ $\gtrsim$ 50 mag according to the standard gas-to-dust ratio by Ryter (1996).  Here it is interesting to mention that all the six Class 0/I sources detected in the IC 1396 N region (distance $\sim$0.75 kpc)  by Getman et al. (2007) having $A_V$ $\gtrsim$ 50 mag show X-ray activity. Non-detection of X-ray emission in probable Class 0/I sources in the NGC 281 region may be due to the detection limit on account of the larger distance. Deeper exposures are required to reach the conclusions.

\section {Summary and conclusions}

In this paper, we present a multiwavelength study of the NGC 281 region using deep wide-field optical $UBVI_c$ data, 
slitless spectroscopy along with the archival data from the surveys such as {\it Chandra}, {\it Spitzer}, 2MASS, IRAS and NVSS.
We made an attempt to  construct a scenario of the global star formation in the NGC 281 complex
by taking into account the effects of massive stars on low-mass star formation. 
The main results from this study are as follow:

\begin{itemize}

\item

The morphology of the central cluster IC 1590 is found to be elongated. The extent of the cluster  is 6.5 pc and the core radius  1.6 pc.  
The maximum age of the ionizing source of the region could be $\sim$ 4 Myr. The minimum reddening $E(B-V)_{min}$ towards the cluster is estimated 
to be 0.32 mag and the cluster shows a small amount of differential reddening ($E(B-V)\sim$ 0.2 mag).  
The distribution of the YSOs selected on the basis of NIR-excess, H$\alpha$ emission and X-ray emission show a  spread 
in the CMD. The age distribution of YSOs indicates a non-coeval star formation in  and around the cluster.

\item 
A slitless spectroscopic survey of the NGC 281 region identifies 12 H$\alpha$ emission stars. Some of these stars are located near the globules/BRCs and show the properties of intermediate-low mass PMS stars. On the basis of NIR excess we identified 87 CTTSs. 
Using a {\it Chandra} archival dataset and NIR colour-colour diagram we also identified 118 WTTSs.
A majority of the identified YSOs (IR-excess, X-ray and H$\alpha$ stars) are low mass PMS stars having ages  $<1-\sim 2$ Myr and masses 0.5-3.5 M$_{\odot}$.

\item
The slope ($\Gamma$) of the MF for the central cluster IC 1590 in the mass range 
$2 < M/M_\odot \le 54$ is found to be $-1.11\pm0.15$. 
The slope of the K-band luminosity function ($0.37\pm0.07$) is similar to the average value ($\sim$0.4) reported for young clusters in the literature (i.e. Lada et al. 1991; Lada \& Lada 1995; Lada \& Lada 2003).

\item 
The distribution of gas and dust obtained from the IRAS, CO and radio continuum maps indicates clumpy structures around  the central cluster. 
The radial distribution, ages and NIR-excesses $\Delta$($H-K$) of the  YSOs as well as the  fraction of CTTSs, suggest triggered  star formation around the cluster.  However we would like to caution the readers that the above statement is not conclusive in view of the scatter in the data. Deeper optical, NIR and MIR observations are needed to have a conclusive view for the  star formation scenario in the region. 
The Class 0/I and Class II sources detected by using the {\it Spitzer} MIR observations indicate that a majority of the Class II sources are X-ray emitting stars, whereas X-ray emission is absent in Class 0/I sources. The spatial distribution of Class 0/I and Class II sources reveals the presence of three sub-clusters in the NGC 281 West region.  The distribution of the Class 0/I and Class II sources in the `a' sub-cluster  indicates
that the Class II sources tend to be located relatively near to the ionizing source.

\end{itemize}

\section*{Acknowledgments}
Authors are thankful to the anonymous referee for useful comments which improved the contents of the paper significantly.
The observations reported in this paper were obtained using the Kiso Schmidt, Japan and the 2 meter HCT at IAO, Hanle, the high altitude station of Indian Institute of Astrophysics. We thank the staff of Kiso Observatory, IAO, Hanle and CREST, Hosakote for their assistance during the observations. This publication makes use of the data from the Two Micron All Sky Survey, which is a joint project of the University of Massachusetts and the Infrared Processing and Analysis Center/California Institute of Technology, funded by the National Aeronautics and Space Administration and the National Science Foundation as well as {\it Chandra} and {\it Spitzer} Data Archives.
AKP and KO acknowledge the financial support given by DST (India) and JSPS (Japan) to carry out the wide field CCD photometry at Kiso. We are also thankful to the Kiso observatory and IAO for allotting the observing time. We thank Annie Robin for letting us use her models of stellar population synthesis.
JB is supported by FONDECYT  No.1080086 and MIDEPLAN ICM Nucleus P07-021-F. 
SS acknowledges the support from Comitee Mixto ESO-GOBIERNO DE CHILE and MIDEPLAN ICM Nucleus P07-021-F.


\begin{table*}
\centering
\caption{\label{log} Log of observations.}
\begin{tabular}{@{}rr@{}}
\hline
Date of observation/Filter& Exp. (sec)$\times$ No. of frames\\
\hline
&Kiso Schmidt telescope, Japan\\
21 November 2004\\
$B$   &  $60\times6,20\times6$\\
$V$   &  $60\times6,10\times6$\\
$I_c$ &  $60\times6,10\times6$\\
27 November 2005\\
$U$   & $180\times6,60\times2$\\
$I_c$ & $10\times2$\\
\\
&Sampurnanand telescope, ARIES\\
07 January 2005\\
$U$   &  $300\times3,120\times1,30\times1$\\
$B$   &  $120\times3,30\times3$\\
$V$   &  $120\times3,30\times3$\\
$I_c$   &  $60\times4,10\times3$\\
\\
&Himalayan {\it Chandra} Telescope, IIA\\
10 October 2005\\
Slitless spectra & $(420\times3)\times4$ \\
Direct Frames  & $(60\times3)\times 4$\\
16 August 2006\\
Slitless spectra & $(300\times2)\times4$ \\
Direct Frames  & $(60\times1)\times 4$\\
\hline
\end{tabular}
\end{table*}

\begin{table*}
\centering
\caption{\label{completness} Completeness Factor (CF) of the optical photometric data in the cluster and field regions.}
\begin{tabular}{@{}rrrrr@{}}
\hline
$V$ range &   \multicolumn{2}{c}{IC 1590}   & Field region \\
(mag)& $r\le2^\prime$ & $ 2^\prime < r\le5^\prime$ & $r\le5^\prime$\\
\hline
  9.5-10.5   &   1.00  &  1.00   & 1.00\\
 10.5-11.5   &   1.00  &  1.00   & 1.00\\
 11.5-12.5   &   1.00  &  1.00   & 1.00\\
 12.5-13.5   &   1.00  &  1.00   & 1.00\\
 13.5-14.5   &   1.00  &  1.00   & 1.00\\
 14.5-15.5   &   1.00  &  0.96   & 0.98\\
 15.5-16.5   &   0.93  &  0.97   & 0.95\\
 16.5-17.5   &   0.74  &  0.94   & 0.96\\
 17.5-18.5   &   0.62  &  0.90   & 0.90\\
 18.5-19.5   &   0.32  &  0.40   & 0.45\\

\hline
\end{tabular}
\end{table*}

\begin{table*}
\centering
\caption{\label{cmpt}Comparison of the present photometry with the available photometry in the literature. 
The difference $\Delta$ (literature-present data) is in magnitude. Mean and $\sigma$ are based on $N$ stars in a $V$ magnitude bin.}
\begin{tabular}{@{}lrrrrrr@{}}
\hline
$V$ range&$\Delta(V)$ && $\Delta(B-V)$&&$\Delta(U-B)$&\\
&($Mean\pm \sigma$)&(N) &($Mean\pm \sigma$)&(N)&( $Mean\pm \sigma$)&(N)\\
\hline

Guetter \& Turner (1997, ccd)\\
 $<12$&$ -0.004\pm0.019$& 5   &$ -0.022\pm0.016$&5  &$ -0.046\pm0.046$&5\\
 12-13&$  0.001\pm0.031$& 5   &$ -0.019\pm0.010$&5  &$ -0.038\pm0.049$&5\\
 13-14&$ -0.011\pm0.022$& 18  &$  0.002\pm0.018$&18 &$ -0.040\pm0.075$&16\\
 14-15&$  0.022\pm0.042$& 16  &$ -0.014\pm0.022$&16 &$  0.010\pm0.086$&14\\
 15-16&$  0.011\pm0.040$&40   &$  0.011\pm0.035$&40 &$  0.051\pm0.102$&21\\
 16-17&$  0.021\pm0.053$&52   &$ -0.026\pm0.042$&50 &$           -   $&-\\
 17-18&$  0.036\pm0.062$&16   &$ -0.002\pm0.077$&16 &$  -            $&-\\
\hline
Guetter \& Turner (1997, pe)\\
 $<12$&$ -0.001\pm0.014$& 5   &$ -0.030\pm0.013$&5  &$ -0.054\pm0.038$&5\\
 12-13&$ -0.018\pm0.011$& 4   &$ -0.016\pm0.014$&4  &$ -0.059\pm0.040$&4\\
 13-14&$  0.015\pm0.011$& 4   &$ -0.013\pm0.014$&4  &$ -0.083\pm0.045$&4\\
\hline
\end{tabular}

ccd: charged coupled device data\\
pe: photo-electric data\\
\end{table*}

\begin{table*}
\centering
\caption{\label{Txray} The  optical, 2MASS and IRAC counterparts of the X-ray sources searched within a match radius of 1 arcsec.
The radial distance is from the cluster center. The complete table is available in the electronic form only. }
\begin{tabular}{@{}cccccccccc@{}}
\hline
Radial Distance& $\alpha_{(2000)}$ & $\delta_{(2000)}$ & $V$ & $(V-I)$&$J$ & $H$ & $K_s$ & 3.6$\mu$m & 4.5$\mu$m \\
 $(^\prime)$ & {\rm $(^h:^m:^s)$} & {\rm $(^o:^\prime:^{\prime\prime)} $} & (mag)& (mag)& (mag)& (mag)& (mag)&(mag) & (mag)\\
\hline
 0.07& 00:52:39.23& +56:37:49.1& 16.095&    1.381& 13.711& 13.035& 12.943&      -&         -  \\
 0.17& 00:52:38.84& +56:37:37.0& 14.391&    1.029& 12.707& 12.229& 12.115&      -&         -  \\
 0.41& 00:52:38.64& +56:38:09.2& 19.155&    1.837& 16.015& 15.501& 15.047&      -&         -  \\
 0.44& 00:52:42.48& +56:37:55.9&      -&        -& 16.060& 15.143& 15.056&      -&         -  \\
 0.54& 00:52:35.82& +56:37:33.6& 17.766&    1.674& 14.829& 13.875& 13.289&      -&         -  \\
  - &     -      &      -     &      -&        -&      -&      -&      -&      -&         -  \\
  - &     -      &      -     &      -&        -&      -&      -&      -&      -&         -  \\
  - &     -      &      -     &      -&        -&      -&      -&      -&      -&         -  \\
\hline
\end{tabular}
\end{table*}

\begin{table*}
\centering
\caption{\label{Tspit} The  2MASS, optical and X-ray counterparts of the IRAC sources searched within a match radius of 1 arcsec.
The radial distance is from the cluster center.  The complete table is available in the electronic form only.  }
\begin{tabular}{@{}ccccccccccc@{}}
\hline
Radial Distance& $\alpha_{(2000)}$ & $\delta_{(2000)}$ & 3.6$\mu$m & 4.5$\mu$m&$J$ & $H$ & $K_s$ & $V$ & $(V-I)$ & X-ray \\
 $(^\prime)$ & {\rm $(^h:^m:^s)$} & {\rm $(^o:^\prime:^{\prime\prime)} $} & (mag)& (mag)& (mag)& (mag)& (mag)&(mag) & (mag) &(Y/N)\\
\hline
 0.95& 00:52:33.90&     +56:37:11.8&  13.821&   13.391& 15.937& 14.981& 14.606& 19.876&  2.169&     N \\
 1.00& 00:52:34.16&     +56:37:04.2&  12.265&   12.171& 13.391& 12.628& 12.366& 16.323&  1.636&     N \\
 1.24& 00:52:31.18&     +56:37:16.5&  15.283&   15.177& 16.090& 15.485& 15.266&      -&      -&     N \\
 1.24& 00:52:34.00&     +56:36:46.3&  14.390&   14.426& 15.233& 14.619& 14.460& 17.617&  1.409&     N \\
 1.25& 00:52:31.32&     +56:37:12.1&  14.461&   14.463& 15.208& 14.566& 14.494& 17.169&  1.217&     N \\
 -  &      -     &           -    &     -  &        -&      -&      -&      -&      -&      -&     - \\
 -  &      -     &           -    &     -  &        -&      -&      -&      -&      -&      -&     - \\
 -  &      -     &           -    &     -  &        -&      -&      -&      -&      -&      -&     - \\
\hline
\end{tabular}
\end{table*}

\begin{table*}
\centering
\caption{\label{iras} Detail of identified cold $IRAS$ point source.}
\begin{tabular}{@{}rrrrrrr@{}}
\hline
$IRAS$ PSC& RA (2000) & Dec. (2000) & $F_{12}$ & $F_{25}$ & $F_{60}$ & $F_{100}$\\
 & {\rm $(^h:^m:^s)$} & {\rm $(^o:^\prime:^{\prime\prime)} $} & (Jy)& (Jy)&(Jy)&(Jy)\\
\hline
  00512+5617 &00:54:14.74 &+56:33:22.7 &   1.98 & 10.84&    44.81 & 218.20\\
\hline
\end{tabular}
\end{table*}

\begin{table*}
\tiny
\centering
\caption{\label{sltt}  The YSOs identified on the basis of H$\alpha$ emission, NIR CCD, MIR CCD, MIR CMD and X-ray emission. The radial distance is from the cluster center. The complete table is available in the electronic form only.   }
\begin{tabular}{@{}lccccccccccccl@{}}
\hline
ID&Radial Dist-& $\alpha_{(2000)}$ & $\delta_{(2000)}$ & $V$ & $(V-I)$&$J$ & $H$ & $K_s$&3.6$\mu$m&4.5$\mu$m & 5.8$\mu$m & 8.0$\mu$m& Remark \\
 & -ance $(^\prime)$ & {\rm $(^h:^m:^s)$} & {\rm $(^o:^\prime:^{\prime\prime)} $} & (mag)& (mag)& (mag)& (mag)& (mag) &(mag)&(mag)&(mag)&(mag)&1,2,3,4,5,6$^a$\\
\hline
1       &  0.07&  00:52:39.22 &  +56:37:49.1& 16.095&   1.381&  13.711&  13.035 & 12.943&       -&       -&       -    &    -   &  3,2  \\
2       &  0.44&  00:52:42.49 &  +56:37:55.9&      -&       -&  16.060&  15.143 & 15.056&       -&       -&       -    &    -   &  3,2  \\
3       &  0.47&  00:52:40.56 &  +56:38:12.1&      -&       -&  15.704&  14.813 & 14.633&       -&       -&       -    &    -   &  3,2  \\
4       &  0.54&  00:52:35.82 &  +56:37:33.6& 17.766&   1.674&  14.829&  13.875 & 13.289&       -&       -&       -    &    -   &  3,1  \\
5       &  0.63&  00:52:35.19 &  +56:37:32.7&      -&       -&  15.674&  14.854 & 14.418&       -&       -&       -    &    -   &  3,2  \\
-       &     -&            - &            -&      -&       -&       -&       - &      -&       -&       -&       -    &    -   &    -  \\
-       &     -&            - &            -&      -&       -&       -&       - &      -&       -&       -&       -    &    -   &    -  \\
-       &     -&            - &            -&      -&       -&       -&       - &      -&       -&       -&       -    &    -   &    -  \\
\hline
\end{tabular}

a: 1=CTTS, 2=WTTS, 3=X-ray, 4=H$\alpha$, 5=Class0/I, 6=ClassII\\
\end{table*}

\begin{table*}
\tiny
\centering
\rotcaption{\label{data_all} MIR, NIR and optical data for Class 0/I, Class II and Class III objects as detected on the basis of MIR  TCD (cf. \S 5).}
\begin{sideways}
\begin{minipage}{15mm}
\begin{tabular}{@{}crrrrrrrrrrrrrcc@{}}
\hline
 ID  &   $\alpha_{2000}$     &  $\delta_{2000}$     &  3.6 $\mu$m & 4.8 $\mu$m & 5.8  $\mu$m  & 8.0  $\mu$m  & $J$     & $H$    & $K_s$    &    $V$ &    $U-B$  &   $B-V$  &    $V-I$ &X-ray&ID(Table \ref{sltt})\\
     &  {\rm $(^h:^m:^s)$} & {\rm $(^o:^\prime:^{\prime\prime)} $}   & (mag) & (mag) &  (mag) &  (mag) &  (mag) &  (mag) &  (mag) &  (mag) &  (mag) &  (mag) &  (mag) &Y/N&\\
\hline

Class 0/I\\
   Ia &00:52:10.99	&+56:30:58.8  &  9.873 &  7.856 &  6.002 &  4.992  &  17.235 & 15.499 & 14.854 &      - &       -  &      - &      - & -&213\\
   Ib &00:52:11.76	&+56:33:04.5  & 10.409 &  9.175 &  8.023 &  7.113  &  17.067 & 15.439 & 13.445 &      - &       -  &      - &      - & -&181\\
   Ic &00:52:29.97	&+56:33:29.2  & 11.129 &  9.372 &  8.114 &  7.114  &       - &      - &      - &      - &       -  &      - &      - & -&123\\
   Id &00:52:17.15	&+56:33:42.5  & 11.678 &  9.538 &  7.937 &  6.846  &       - &      - &      - &      - &       -  &      - &      - & -&137\\
   Ie &00:52:26.58	&+56:33:25.5  & 11.845 & 10.677 &  9.793 &  9.170  &  17.614 & 17.246 & 14.904 &      - &       -  &      - &      - & -&127\\
   If &00:52:24.56	&+56:33:50.1  & 12.175 & 10.345 &  9.668 &  9.714  &       - &      - &      - &      - &       -  &      - &      - & -&121\\
   Ig &00:52:13.48	&+56:33:41.4  & 12.232 & 11.233 & 10.323 &  9.459  &  18.614 & 16.133 & 14.317 &      - &       -  &      - &      - & -&156\\
   Ih &00:52:16.45	&+56:31:45.9  & 12.765 & 12.256 & 11.478 & 10.362  &  16.519 & 14.960 & 13.984 &      - &       -  &      - &      - & -&200\\
   Ii &00:52:15.92	&+56:33:49.5  & 13.065 & 11.618 & 10.383 &  9.644  &  18.744 & 16.025 & 15.401 &      - &       -  &      - &      - & -&139\\
   Ij &00:52:19.56	&+56:32:58.4  & 13.255 & 12.723 & 12.101 & 10.955  &  15.962 & 14.941 & 14.371 &      - &       -  &      - &      - & -&164\\
   Ik &00:52:32.54	&+56:32:33.6  & 13.401 & 12.822 & 11.557 &  9.870  &  16.697 & 15.463 & 15.115 & 19.743 &       -  &      - &  1.758 & -&149\\
   Il &00:52:21.37	&+56:30:40.4  & 13.711 & 12.544 & 11.454 & 10.546  &       - &      - &      - &      - &       -  &      - &      - & -&209\\
   Im &00:52:25.55	&+56:33:32.6  & 15.221 & 13.185 & 11.631 & 10.246  &       - &      - &      - &      - &       -  &      - &      - & -&126\\
   In &00:52:31.37	&+56:33:31.2  & 15.625 & 13.613 & 12.470 & 11.700  &       - &      - &      - &      - &       -  &      - &      - & -&120\\\\

Class II\\
  IIa &00:52:20.84	&+56:33:08.2  & 10.048 &  9.305 &  8.535 &  7.695  &  16.288 & 14.617 & 12.502 &      - &       -  &      - &      - & Y&147\\
  IIb &00:52:22.39	&+56:34:26.5  & 10.957 & 10.422 &  9.871 &  8.993  &  14.741 & 13.121 & 12.207 &      - &       -  &      - &      - & Y&101\\
  IIc &00:52:12.49	&+56:34:13.7  & 11.148 & 10.449 &  9.714 &  8.798  &  17.305 & 14.889 & 13.359 &      - &       -  &      - &      - & -&141\\
  IId &00:52:27.34	&+56:34:03.2  & 11.345 & 10.925 & 10.410 &  9.629  &  15.891 & 14.148 & 12.973 &      - &       -  &      - &      - & Y&102\\
  IIe &00:52:27.31	&+56:32:04.8  & 11.375 & 10.583 &  9.867 &  9.125  &  15.832 & 14.231 & 13.388 &      - &       -  &      - &      - & -&179\\
  IIf &00:52:35.75	&+56:34:31.1  & 11.391 & 10.820 & 10.330 &  9.896  &  15.383 & 13.871 & 12.827 &      - &       -  &      - &      - & -&71 \\
  IIg &00:52:18.83	&+56:33:05.9  & 11.590 & 11.187 & 10.813 & 10.385  &  14.510 & 13.390 & 12.821 &      - &       -  &      - &      - & Y&159\\
  IIh &00:52:19.31	&+56:31:14.4  & 11.814 & 11.269 & 10.714 &  9.967  &  14.582 & 13.479 & 12.947 & 19.256 &       -  &      - &  2.708 & Y&204\\
  IIi &00:52:21.59	&+56:31:31.8  & 11.947 & 11.517 & 11.108 & 10.427  &  15.087 & 13.769 & 13.031 &      - &       -  &      - &      - & Y&198\\
  IIj &00:52:10.89	&+56:34:06.3  & 12.170 & 11.634 & 11.212 & 10.561  &  17.328 & 15.770 & 14.193 &      - &       -  &      - &      - & -&153\\
  IIk &00:52:33.69	&+56:35:59.8  & 13.010 & 12.775 & 12.496 & 11.841  &  14.976 & 14.000 & 13.657 & 17.963 &       -  &  1.177 &  1.904 & Y&29 \\
  IIl &00:52:19.49	&+56:32:52.7  & 13.169 & 12.694 & 12.028 & 11.484  &  16.189 & 15.358 & 14.957 &      - &       -  &      - &      - & -&170\\
  IIm &00:52:14.89	&+56:31:44.8  & 13.885 & 13.340 & 12.702 & 11.638  &  17.147 & 15.553 & 15.261 &      - &       -  &      - &      - & -&202\\\\

Class III\\
 IIIa &00:52:16.82	&+56:31:33.4  &  8.601 &  8.477 &  8.431 &  8.486  &   8.736 &  8.556 &  8.515 &      - &       -  &      - &      - & Y&- \\
 IIIb &00:52:10.31	&+56:31:31.6  &  9.252 &  8.835 &  8.513 &  8.192  &  10.904 & 10.402 &  9.974 & 12.828 &   0.027  &  0.893 &  1.362 & -&207\\
 IIIc &00:52:14.65	&+56:34:39.6  &  9.699 &  9.794 &  9.634 &  9.569  &  10.558 & 10.000 &  9.839 & 12.682 &   0.790  &  1.145 &  1.268 & -&- \\
 IIId &00:52:34.26	&+56:32:21.0  & 10.173 & 10.230 & 10.200 & 10.268  &  11.137 & 10.449 & 10.287 & 13.594 &   1.090  &  1.352 &  1.448 & -&- \\
 IIIe &00:52:36.28	&+56:36:00.0  & 10.271 & 10.315 & 10.207 &  9.944  &  10.395 & 10.313 & 10.334 & 10.867 &  -0.507  &  0.225 &  0.362 & -&- \\
 IIIf &00:52:42.88	&+56:35:16.3  & 10.712 & 10.727 & 10.642 & 10.789  &  10.997 & 10.835 & 10.768 & 11.885 &   0.240  &  0.467 &  0.572 & -&- \\
 IIIg &00:52:40.55	&+56:35:50.1  & 11.224 & 11.236 & 11.111 & 11.205  &  11.859 & 11.548 & 11.391 & 13.517 &   0.549  &  0.658 &  0.933 & Y&- \\
 IIIh &00:52:25.71	&+56:34:15.5  & 11.913 & 11.300 & 10.783 & 10.523  &  17.138 & 14.700 & 13.611 &      - &       -  &      - &      - & Y&95\\
\hline
\end{tabular}
\end{minipage}
\end{sideways}
\end{table*}

\begin{table*}
\centering
\caption{\label{Tage}  The mass and age of the YSOs having optical counterparts along with the associated errors.
The ID is as same as in Table \ref{sltt}.  }
\begin{tabular}{@{}rrr|rrr|rrr@{}}
\hline
ID&  Mass $\pm \sigma$& Age $\pm \sigma$&ID&  Mass $\pm \sigma$& Age $\pm \sigma$&ID&  Mass $\pm \sigma$& Age $\pm \sigma$ \\
  &  (M$_\odot$)   &  (Myr)&&  (M$_\odot$)   &  (Myr)&&  (M$_\odot$)   &  (Myr) \\
\hline
     1& $2.20\pm0.11$ &$1.01\pm0.21$&     48& $1.15\pm0.07$ &$0.58\pm0.09$&     192& $0.35\pm0.02$ &$0.37\pm0.27$ \\
     4& $1.11\pm0.08$ &$1.36\pm0.30$&     50& $0.52\pm0.03$ &$0.72\pm0.06$&     197& $0.53\pm0.04$ &$0.42\pm0.16$ \\
     6& $0.84\pm0.05$ &$0.68\pm0.09$&     51& $0.57\pm0.05$ &$1.89\pm0.51$&     199& $0.96\pm0.07$ &$0.34\pm0.04$ \\
     7& $2.72\pm0.13$ &$3.09\pm0.39$&     53& $0.89\pm0.07$ &$2.19\pm0.55$&     201& $0.45\pm0.03$ &$1.01\pm0.12$ \\
     8& $0.53\pm0.05$ &$2.10\pm0.61$&     59& $0.36\pm0.01$ &$0.31\pm0.23$&     206& $0.34\pm0.01$ &$0.10\pm0.01$ \\
     9& $6.29\pm0.30$ &$0.36\pm0.04$&     60& $1.08\pm0.06$ &$0.43\pm0.05$&     209& $5.34\pm0.07$ &$0.17\pm0.02$ \\
    12& $1.84\pm0.10$ &$1.19\pm0.24$&     63& $0.54\pm0.04$ &$1.45\pm0.34$&     214& $0.81\pm0.06$ &$3.10\pm0.71$ \\
    14& $0.72\pm0.07$ &$0.67\pm0.10$&     65& $0.82\pm0.04$ &$0.92\pm0.12$&     217& $0.44\pm0.01$ &$0.10\pm0.00$ \\
    18& $0.97\pm0.08$ &$3.71\pm0.97$&     67& $0.91\pm0.06$ &$3.96\pm0.82$&     220& $1.48\pm0.03$ &$4.86\pm0.29$ \\
    22& $0.65\pm0.05$ &$2.20\pm0.58$&     70& $1.87\pm0.05$ &$2.79\pm0.57$&     221& $1.70\pm0.04$ &$3.82\pm0.68$ \\
    23& $0.73\pm0.07$ &$1.69\pm0.40$&     75& $0.53\pm0.03$ &$1.01\pm0.14$&     227& $1.43\pm0.03$ &$4.85\pm0.27$ \\
    25& $3.31\pm0.14$ &$1.66\pm0.27$&     76& $1.09\pm0.06$ &$0.77\pm0.12$&     228& $0.94\pm0.03$ &$5.00\pm0.00$ \\
    27& $1.66\pm0.08$ &$1.91\pm0.40$&     79& $0.79\pm0.03$ &$5.00\pm0.00$&     229& $3.10\pm0.09$ &$0.96\pm0.25$ \\
    29& $0.75\pm0.05$ &$0.77\pm0.07$&     80& $1.11\pm0.07$ &$2.67\pm0.62$&     230& $0.59\pm0.04$ &$1.38\pm0.27$ \\
    30& $0.84\pm0.06$ &$1.59\pm0.34$&     85& $1.14\pm0.03$ &$4.99\pm0.03$&     231& $0.82\pm0.04$ &$0.41\pm0.03$ \\
    31& $0.49\pm0.03$ &$0.87\pm0.09$&     87& $1.09\pm0.07$ &$2.37\pm0.53$&     233& $1.22\pm0.07$ &$0.39\pm0.04$ \\
    32& $0.92\pm0.06$ &$0.85\pm0.15$&     88& $0.73\pm0.05$ &$1.24\pm0.18$&     234& $0.87\pm0.03$ &$5.00\pm0.00$ \\
    34& $0.51\pm0.03$ &$0.78\pm0.07$&     90& $0.87\pm0.05$ &$1.67\pm0.34$&     235& $0.64\pm0.04$ &$4.48\pm0.72$ \\
    36& $0.60\pm0.04$ &$0.80\pm0.15$&     96& $1.39\pm0.05$ &$2.74\pm0.55$&     236& $0.52\pm0.02$ &$0.10\pm0.00$ \\
    38& $1.64\pm0.04$ &$3.41\pm0.65$&    108& $0.79\pm0.05$ &$1.68\pm0.29$&     242& $0.40\pm0.02$ &$0.91\pm0.08$ \\
    40& $0.49\pm0.01$ &$0.10\pm0.01$&    125& $0.88\pm0.05$ &$0.85\pm0.14$&     245& $0.49\pm0.04$ &$1.12\pm0.18$ \\
    43& $0.81\pm0.04$ &$0.95\pm0.13$&    149& $0.78\pm0.03$ &$5.00\pm0.01$&     250& $0.35\pm0.01$ &$0.30\pm0.23$ \\
    44& $1.34\pm0.03$ &$4.96\pm0.05$&    150& $0.78\pm0.03$ &$5.00\pm0.01$&     259& $1.03\pm0.06$ &$0.97\pm0.16$ \\
    45& $0.90\pm0.05$ &$4.39\pm0.72$&    169& $1.04\pm0.03$ &$5.00\pm0.00$&     268& $0.82\pm0.03$ &$5.00\pm0.00$ \\
    46& $0.73\pm0.05$ &$1.61\pm0.27$&    175& $0.30\pm0.01$ &$0.12\pm0.06$&                                       \\
    47& $0.48\pm0.03$ &$0.57\pm0.11$&    187& $1.32\pm0.03$ &$4.97\pm0.04$&                                       \\
\hline
\end{tabular}
\end{table*}

\begin{table*}
\centering
\caption{\label{statT} Statistics of probable CTTSs in three sub-regions. Numbers given in parentheses are in
percentage. }
\begin{tabular} {r r c c c l l}
\hline
Region    &  Total  &        NIR-excess  & $H\alpha$    &    Probable CTTSs & $\Delta (H-K)$  & $A_V$    \\
          &  stars  &        stars       & stars      & (NIR-excess  &(mag)&  (mag) \\
          &         &                    &            &  + H$\alpha$ stars) & &\\
\hline
stars with error less than 0.1 mag\\
\hline
  1       &  168    &         2 (2.5)     &  3 (3.8)   &      5 (6.3)     &$0.06\pm0.03$ & $1.6\pm0.2$ \\
  2       &  107    &         6 (33.3)    &  1 (5.5)   &      7 (38.9)    &$0.08\pm0.04$ & $3.9\pm1.9$ \\
  3       &  97     &         4 (50.0)    &  3 (37.5)  &      6 (75.0)$^a$&$0.16\pm0.05$ & $3.6\pm2.3$  \\
Field     &   89    &         $-$         &     $-$       &    $-$    & $-$ & $-$      \\
\hline
all stars independent of errors\\
\hline
  1       &  425    &         8 (3.3)     &  3 (1.3)   &      11 (4.6)     &$0.12\pm0.08$ & $2.5\pm1.8$ \\
  2       &  313    &         32 (25.2)   &  1 (0.7)   &      33 (26.0)    &$0.10\pm0.07$ & $4.4\pm2.8$ \\
  3       &  241    &         12 (21.8)   &  3 (5.5)   &      14 (25.5)$^a$&$0.11\pm0.07$ & $4.6\pm2.8$  \\
Field     &   186   &         $-$         &     $-$       &    $-$    & $-$ & $-$      \\
\hline
\end{tabular}

a: One $H\alpha$ stars have NIR-excess\\
\end{table*}

\begin{table*}
\centering
\caption{\label{ctts} CTTS fraction $f_{CTTS} = N_{CTTS}/(N_{CTTS} + N_{WTTS})$.}
\begin{tabular}{@{}ccccccc@{}}
\hline
Region &   \multicolumn{2}{c}{CTTS} &   \multicolumn{2}{c}{WTTS}&   \multicolumn{2}{c}{$f_{CTTS}$}   \\
       & $error \leq 0.1$ & all & $error \leq 0.1$ & all & $error \leq 0.1$ & all \\

\hline
1 & 2 &  8 & 38& 56& 0.05&0.12\\
2 & 6 & 32 & 18& 39& 0.25&0.45\\

\hline

\end{tabular}
\end{table*}

\clearpage


\newpage

\clearpage

\end{document}